\documentclass[12pt, draftclsnofoot, onecolumn, romanappendices]{IEEEtran}

\IEEEoverridecommandlockouts

\usepackage{cite}
\usepackage{graphicx}
\usepackage[cmex10]{amsmath}
\usepackage{amssymb}
\usepackage{array}
\usepackage{psfrag}
\usepackage{arydshln}
\usepackage{pifont}
\usepackage{cancel}
\usepackage{wasysym}
\usepackage{color}

\newcommand{\cmark}{\ding{51}}
\newcommand{\xmark}{\ding{55}}

\usepackage{multirow}
\usepackage[pdftex,hypertexnames=false]{hyperref}

\usepackage{footnote}
\makesavenoteenv{tabular}

\def\baselinestretch{1.42}

\newtheorem{lemma}{Lemma}
\newtheorem{theorem}{Theorem}
\newtheorem{corollary}{Corollary}

\newtheorem{remark}{Remark}

\newcommand{\mb}{\mathbf}

\newcommand{\uh}{{\mb{h}}}
\newcommand{\uha}{{\uh_1}}

\newcommand{\uhaH}{\uh_1^{\cal{H}}}

\newcommand{\uhda}{{\mb{h}_{\text{d},1}}}

\newcommand{\uhra}{{\mb{h}_{\text{r},1}}}

\newcommand{\ug}{{\mb{g}}}

\newcommand{\umu}{\pmb{\mu}}

\newcommand{\uTheta}{{\pmb{\Theta}}}
\newcommand{\uPsi}{{\pmb{\Psi}}}

\newcommand{\uLambda}{{\pmb{\Lambda}}}
\newcommand{\ulambda}{{\pmb{\lambda}}}
\newcommand{\uSigma}{{\pmb{\Sigma}}}

\newcommand{\usigma}{{\pmb{\sigma}}}

\newcommand{\ur}{{\mb{r}}}

\newcommand{\uA}{{\mb{A}}}
\newcommand{\uAaa}{{\mb{A}}_{11}}
\newcommand{\uAab}{{\mb{A}}_{12}}
\newcommand{\uAba}{{\mb{A}}_{21}}
\newcommand{\uAbb}{{\mb{A}}_{22}}

\newcommand{\uAinv}{{\mb{A}}^{-1}}
\newcommand{\uAinvaa}{{\mb{A}}^{11}}
\newcommand{\uAinvab}{{\mb{A}}^{12}}
\newcommand{\uAinvba}{{\mb{A}}^{21}} 
\newcommand{\uAinvbb}{{\mb{A}}^{22}}

\newcommand{\uAH}{{\mb{A}}^{\cal{H}}}
\newcommand{\uAinvH}{{{\mb{A}}^{-{\cal{H}}}}}
\newcommand{\uQ}{{\mb{Q}}}

\newcommand{\uD}{{\mb{D}}}

\newcommand{\uB}{{\mb{B}}}
\newcommand{\uBH}{{\mb{B}}^{\cal{H}}}
\newcommand{\uBtilde}{\widetilde{{\mb{B}}}}
\newcommand{\uBtildeH}{{\uBtilde}^{\cal{H}}}

\newcommand{\uM}{{\mb{M}}}
\newcommand{\uMH}{{\mb{M}}^{\cal{H}}}
\newcommand{\uH}{{\mb{H}}}
\newcommand{\uHH}{{\mb{H}}^{\cal{H}}}
\newcommand{\uHhat}{\widehat{\uH}}

\newcommand{\uHa}{{\mb{H}_1}}
\newcommand{\uHaH}{{\mb{H}_1^{\cal{H}}}}
\newcommand{\uHb}{{\mb{H}_2}}
\newcommand{\uHbH}{{\mb{H}_2^{\cal{H}}}}

\newcommand{\uHw}{{{\mb{H}}}_{\text{w}}}
\newcommand{\uHwa}{{{\mb{H}}}_{\text{w},1}}
\newcommand{\uHwb}{{{\mb{H}}}_{\text{w},2}}

\newcommand{\uHd}{{\uH_{\text{d}}}}
\newcommand{\uHda}{{\uH_{\text{d},1}}}
\newcommand{\uHdatilde}{{\widetilde{\uH}_{\text{d},1}}}
\newcommand{\uHdb}{{\uH_{\text{d},2}}}
\newcommand{\uHdaH}{{\uH_{\text{d},1}^{\cal{H}}}}
\newcommand{\uHdatildeH}{{\widetilde{\uH}_{\text{d},1}^{\cal{H}}}}
\newcommand{\uHdbH}{{\uH_{\text{d},2}^{\cal{H}}}}
\newcommand{\uHdH}{{\uH_{\text{d}}^{\cal{H}}}}

\newcommand{\uHr}{{\uH_{\text{r}}}}

\newcommand{\uHrH}{{\uH_{\text{r}}^{\cal{H}}}}

\newcommand{\uHra}{{\uH_{\text{r},1}}}
\newcommand{\uHrb}{{\uH_{\text{r},2}}}

\newcommand{\uHrbH}{{\uH_{\text{r},2}^{\cal{H}}}}

\newcommand{\NT}{{N_{\text{T}}}}

\newcommand{\NR}{{N_{\text{R}}}}
\newcommand{\uP}{{\mb{P}}}
\newcommand{\uR}{{\mb{R}}}

\newcommand{\uRT}{{\mathbf{R}_{\text{T}}}}
\newcommand{\uRTH}{{\mathbf{R}_{\text{T}}^{\cal{H}}}}

\newcommand{\uRTK}{{\mathbf{R}_{\text{T,}K}}}
\newcommand{\uRTKH}{{\mathbf{R}^{\cal{H}}_{\text{T,}K}}}
\newcommand{\uRTKhat}{{\widehat{\mbR}_{\text{T},K}}}

\newcommand{\uRTKinv}{{\mathbf{R}_{\text{T,}K}^{-1}}}
\newcommand{\uRTKinvhat}{\widehat{\mbR}_{\text{T},K}^{-1}}
\newcommand{\uRTsqrt}{{\mathbf{R}_{\text{T}}^{1/2}}}

\newcommand{\uX}{{\mb{X}}}
\newcommand{\uXH}{{\mb{X}}^{\cal{H}}}

\newcommand{\uS}{{\mb{S}}}

\newcommand{\ux}{{\mb{x}}}

\newcommand{\uy}{{\mb{y}}}

\newcommand{\un}{{\mb{n}}}

\newcommand{\untilde}{{\widetilde{\mb{n}}}}

\newcommand{\mzero}{\mb{0}}

\newcommand{\mbR}{{\mb{R}}}

\newcommand{\uF}{{\mb{F}}}
\newcommand{\uFH}{{\mb{F}}^{\cal{H}}}

\newcommand{\uU}{{\mb{U}}}

\newcommand{\uUH}{{\mb{U}}^{\cal{H}}}
\newcommand{\uV}{{\mb{V}}}
\newcommand{\uVH}{{\mb{V}}^{\cal{H}}}
\newcommand{\uW}{{\mb{W}}}
\newcommand{\uWaa}{{\mb{W}}_{11}}
\newcommand{\uWab}{{\mb{W}}_{12}}
\newcommand{\uWba}{{\mb{W}}_{21}}
\newcommand{\uWbb}{{\mb{W}}_{22}}
\newcommand{\uWinv}{{\uW}^{-1}}
\newcommand{\uWinvaa}{{\mb{W}}^{11}}
\newcommand{\uWinvab}{{\mb{W}}^{12}}
\newcommand{\uWinvba}{{\mb{W}}^{21}}
\newcommand{\uWinvbb}{{\mb{W}}^{22}}

\newcommand{\uWhat}{\widehat{\uW}}

\newcommand{\uWhatinv}{\uWhat^{-1}}

\newcommand{\uGamma}{{\pmb{\Gamma}}}

\newcommand{\mbI}{\mb{I}}

\newcommand{\nid}{{d}}

\newcommand{\trace}{\mbox{tr}}

\newcommand{\RT}{\mathbf{R}_{\text{T}}}

\newcommand{\RTaa}{\mathbf{R}_{{\text{T},K}_{11}}}
\newcommand{\RTupaa}{\mathbf{R}_{{\text{T},K}}^{11}}

\newcommand{\RTbb}{\mathbf{R}_{{\text{T},K}_{22}}}

\newcommand{\RTupbb}{\mathbf{R}_{{\text{T},K}}^{22}}
\newcommand{\RTbbinva}{\mathbf{R}_{{\text{T},K}_{22}}^{-1}}

\newcommand{\RTab}{\mathbf{R}_{{\text{T},K}_{12}}}
\newcommand{\RTabH}{\mathbf{R}_{{\text{T},K}_{12}}^{\cal{H}}}
\newcommand{\RTupab}{\mathbf{R}_{{\text{T},K}}^{12}}

\newcommand{\RTba}{\mathbf{R}_{{\text{T},K}_{21}}}

\newcommand{\RTupba}{\mathbf{R}_{{\text{T},K}}^{21}}

\newcommand{\RThat}{\widehat{\mbR}_{\text{T},K}}
\newcommand{\RThatinv}{\widehat{\mbR}_{\text{T},K}^{-1}}

\newcommand{\RThatupaa}{\widehat{\mathbf{R}}_{{\text{T},K}}^{11}}

\title{Schur Complement Based Analysis \\of MIMO Zero-Forcing for Rician Fading}

\author{
Constantin Siriteanu\thanks{C.~Siriteanu is with the Dept. of Information Systems Engineering, Osaka University, Japan.}, Akimichi Takemura\thanks{A. Takemura is with the Dept. of Mathematical Informatics, University of Tokyo, Japan.},
Satoshi Kuriki\thanks{S. Kuriki is with the Institute of Statistical Mathematics, Tokyo,
Japan.}, \\
Donald St. P. Richards\thanks{D. St.~P.~Richards is with the Dept. of Statistics, Pennsylvania State University, University Park, Pennsylvania, U.S.A.},
Hyundong Shin\thanks{H.~Shin is with the Dept. of Electronics and Radio Engineering, Kyung Hee University, South Korea.}}

\begin{document}

\maketitle

\begin{abstract}
For multiple-input/multiple-output (MIMO) spatial multiplexing with zero-forcing detection (ZF), signal-to-noise ratio (SNR) analysis for Rician fading involves the cumbersome noncentral-Wishart distribution (NCWD) of the transmit sample-correlation (Gramian) matrix.
An \textsl{approximation} with a \textsl{virtual} CWD previously yielded for the ZF SNR an approximate (virtual) Gamma distribution.
However, analytical conditions qualifying the accuracy of the SNR-distribution approximation were unknown.
Therefore, we have been attempting to exactly characterize ZF SNR for Rician fading.
Our previous attempts succeeded only for the sole Rician-fading stream under Rician--Rayleigh fading, by writing it as scalar Schur complement (SC) in the Gramian.
Herein, we pursue a more general, matrix-SC-based analysis to characterize SNRs when several streams may undergo Rician fading.
On one hand, for full-Rician fading, the SC distribution is found to be exactly a CWD if and only if a channel-mean--correlation \textsl{condition} holds. 
Interestingly, this CWD then coincides with the \textsl{virtual} CWD ensuing from the \textsl{approximation}.
Thus, under the \textsl{condition}, the actual and virtual SNR-distributions coincide.
On the other hand, for Rician--Rayleigh fading, the matrix-SC distribution is characterized in terms of determinant of matrix with elementary-function entries, which also yields a new characterization of the ZF SNR.
Average error probability results validate our analysis vs.~simulation.
\end{abstract}



\section{Introduction}
\label{section_introduction}

\subsection{Background, Motivation, Scope, and Main Assumptions}
\label{section_background}

Multiple-input/multiple-output (MIMO) communications principles have maintained substantial research interest\cite{marzetta_tit_99}\cite{paulraj_book_05}\cite{gesbert_spm_07}\cite{jung_jcn_13}\cite{Matthaiou_jsac_13} and have also been adopted in standards\cite{lee_09_jwcn}\cite{li_cm_10}.
However, gaps remain in our ability to evaluate MIMO performance, based on analysis, for realistic channel propagation conditions and relatively simple transceiver processing: e.g., for MIMO spatial-multiplexing for Rician fading and linear detection methods, such as zero-forcing detection (ZF)\cite{kiessling_spawc_03}\cite{siriteanu_tvt_11} or minimum mean-square-error detection (MMSE)\cite{mckay_tcomm_09}.

Rician fading is both theoretically more general and practically more realistic than Rayleigh fading (which yields simpler analysis), according to the state-of-the-art WINNER II channel model\cite{winner_d_1_1_2_v_1_2}.
ZF has relatively-low implementation complexity, and, thus, is attractive for practical implementation, as recently acknowledged under the massive-MIMO framework\cite{jung_jcn_13}\cite{Matthaiou_jsac_13}.

Herein, we study MIMO ZF\footnote{A study of MMSE is left for future work.} under Rician and Rayleigh fading conditions and mixtures that (beside promoting analysis tractability) may occur in macrocells, microcells, and heterogeneous networks, as discussed in\cite{mckay_tcomm_09}\cite{siriteanu_twc_13} and relevant references therein. 


We consider a MIMO system whereby the symbol streams transmitted from $ \NT $ antennas are received with $ \NR \ge \NT $ antennas. 
The $ \NR \times \NT $ channel matrix $ \uH $ is assumed Gaussian-distributed.
For analysis tractability, we assume that elements on different rows of $ \uH $ are uncorrelated, and that each of the $ \NR $ rows of $ \uH $ has the same covariance matrix.
Then, given the transmit sample-correlation matrix $ \uW = \uHH \uH $, also known as Gramian matrix\cite[p.~288]{gentle_book_07}, the ZF signal-to-noise ratio (SNR) for a stream is determined by the corresponding diagonal element of $ \uWinv $\cite[Eq.~(5)]{gore_cl_02}.

\subsection{Previous Work on MIMO ZF for Rician Fading}
\label{section_previous_work}

For MIMO ZF under Rayleigh-only fading, the stream SNRs have been shown to be Gamma-distributed in\cite{gore_cl_02}, based on the fact that, when the mean $ \uHd $ of $ \uH $ is zero, $ \uW $ has a central-Wishart distribution\footnote{For simplicity, N/CWD stand herein for both `non/central-Wishart distribution' and `non/central-Wishart-distributed'.} (CWD)\cite{mckay_tit_05}, and then $ \uWinv $ has a known inverse-CWD\cite[p.~97]{muirhead_book_05}.

On the other hand, under Rician fading, i.e., when $ \uHd \neq \mzero $, $ \uW $ is NCWD\cite{mckay_tit_05}\cite{kuriki_aism_10}, and then $ \uWinv $ has an unknown distribution.
Therefore, for MIMO ZF under full-Rician fading\footnote{I.e., all streams undergo Rician fading.}, we attempted in\cite{siriteanu_tvt_11} to characterize the ZF SNR distribution by approximating the actual NCWD of the  Gramian matrix $ \uW $ with a \textsl{virtual} CWD of equal mean.
This approximation had been proposed in\cite{steyn_roux_sasj_72} and had been exploited for MIMO ZF analysis several times\cite[Refs.~24-27,30,31]{siriteanu_tvt_11}, because it yields a simple virtual Gamma distribution to approximate the unknown actual distribution of the ZF SNR.

However, the accuracy of this SNR-distribution approximation has been qualified only empirically.
Thus, numerical results shown without explanation or support mostly for $ \uHd $ of rank $ r = 1 $ obtained as outer-product of receive and transmit array-steering vectors in\cite[Refs.~24-27,30,31]{siriteanu_tvt_11} found the approximation reliable.
In\cite{siriteanu_tvt_11}, we also found it most accurate for such rank-1 $ \uHd $; higher $ r $ yielded poorer accuracy, and $ r = \NT $ made the approximation unusable. 
However, for $ r = 1 $ we also found that different mean--correlation combinations yield different accuracies.

Because\cite{siriteanu_tvt_11} found the ZF SNR distribution approximation not consistently reliable and because analytical conditions for its accuracy were unknown, we pursued in\cite{siriteanu_twc_13} an exact ZF-SNR analysis. 
That analysis was found tractable only for the case when the intended Stream 1 undergoes Rician fading whereas interfering Streams $ 2,\cdots,\NT $ undergo Rayleigh fading, i.e., Rician$(1)$/Rayleigh$(\NT-1)$ fading, for which, incidentally, $ r = 1 $. 
Proceeding in\cite{siriteanu_twc_13} from the \textsl{vector}--\textsl{matrix} partitioning according to fading types $ \uH = \begin{matrix} ( {\uh_1} & {\uH_2})  \end{matrix} $, we could write the Stream-1 ZF SNR in terms of the scalar Schur complement (SC)\cite{zhang_book_05}\cite{ouellette1_laa_81}\cite[Sec.~3.4]{gentle_book_07} of submatrix $\uW_{22} = \uHbH \uHb $ in the NCWD Gramian matrix $ \uW $.

Note that the SC arises ``naturally'' in statistical analyses as the (sample) correlation matrix of the conditioned Gaussian distribution\cite[p.~186]{zhang_book_05}, as also exemplified in\cite{siriteanu_twc_13} and this paper.
Also, after minimizing a Hermitian form over some variables, the matrix in the ensuing Hermitian form is a SC\cite[Eqs.~(A.13-14), p.~650]{boyd_book_04}.

In\cite{siriteanu_twc_13}, we recast the scalar SC as a Hermitian form whereby the vector and matrix correspond, respectively, to the intended and interfering streams\cite[Eq.~(9)]{siriteanu_twc_13}\cite[Eq.~(7)]{kiessling_spawc_03}.
By first conditioning on and then averaging over $ \uHb $, we expressed exactly, in\cite[Eq.~(31)]{siriteanu_twc_13}, the moment generating function (m.g.f.) of the ZF SNR for the Rician-fading Stream 1, in terms of the confluent hypergeometric function $ {_1\!F_1} \left(N; \NR; \sigma_1 \right) $, where $ N = \NR - \NT + 1 $, and scalar $ \sigma_1 $ is a function of channel mean and transmit-correlation\footnote{This SNR m.g.f.~was then written an infinite linear combination of m.g.f.s of Gamma distributions in\cite[Eq.~(37)]{siriteanu_twc_13}.}.

Finally, average error probability (AEP) results shown in\cite[Figs.~1, 2]{siriteanu_twc_13} for Rician$(1)$/ Rayleigh$(\NT-1)$ fading further supported our conclusion from\cite{siriteanu_tvt_11} that the actual--virtual SNR-distribution approximation is inconsistently accurate even for $ r = 1 $.
On the other hand, results shown in\cite[Fig.~10]{siriteanu_twc_13} for Rayleigh$(1)$/Rician$(\NT-1)$ fading suggested that the approximation can be very accurate even for $ r > 1 $.
The empirical observations from\cite{siriteanu_tvt_11}\cite{siriteanu_twc_13} have prompted us to seek the analytical condition that renders this approximation exact.


\subsection{Goals and Approach}
\label{section_goals_approach}

Herein, we explore and exploit the relationship between the matrix-SC and ZF SNRs to statistically characterize the latter when several streams may experience Rician fading.
Also, we aim to reveal the necessary and sufficient \textsl{condition} for the matrix-SC to become CWD, and for the virtual Gamma distribution to become the exact distribution of ZF SNRs.

Thus, based on the matrix--matrix partitioning $ \uH = ( \begin{matrix} \uHa & \uHb \end{matrix} )$, where $ \uHa $ is $ \NR \times v $, we characterize the distribution of the ensuing $ v \times v $ matrix-SC of $\uW_{22} $ in $ \uW $, denoted $ \uGamma_1 $. 
This helps characterize the ZF SNR distributions of streams corresponding to $ \uHa $ when several streams may undergo Rician fading.
It also helps reveal the sought \textsl{condition}.


\subsection{Contributions}
\label{section_Contributions}

First, for full-Rician fading, we show that $ \uGamma_1 $ conditioned on $ \uHb $ is NCWD, and state the necessary and sufficient \textsl{condition} --- found to be a special relationship between the means and column-correlations of $ \uHa $ and $ \uHb $ --- that yields a CWD for the unconditioned $ \uGamma_1 $, and Gamma distributions for ZF SNRs.
Then, we prove that the actual and virtual CWDs for the matrix-SC $ \uGamma_1 $ coincide under the \textsl{condition}.
Consequently, the actual (generally unknown) and virtual (Gamma) distributions of the ZF SNRs for the streams corresponding to $ \uHa $ also coincide.
Thus, surprisingly, although these streams may undergo Rician fading, their SNRs are distributed as when they undergo Rayleigh fading, which has not been known possible.
Importantly, this \textsl{condition} qualifies analytically, for the first time, the relationship between the actual distribution of the ZF SNR under Rician fading and the virtual Gamma distribution. 
Thus, it helps corroborate approximation-accuracy observations from\cite{siriteanu_tvt_11}\cite[Figs.~1,2,10]{siriteanu_twc_13}.
Then, as it is unrelated to condition $ r = 1 $ imposed in\cite[Refs.~24-27,30,31]{siriteanu_tvt_11}, it also explains the inconsistent approximation accuracy found for $ r = 1 $ in\cite{siriteanu_tvt_11}.

Second, we characterize exactly the distribution of the matrix-SC $ \uGamma_1 $ for zero-mean $ \uHb $, i.e., for Rician$(v)$/Rayleigh$(\NT-v)$ fading.
The m.g.f.~of $ \uGamma_1 $ is deduced in terms of the hypergeometric function $ {{_0\!F_0}} (\uS, \uLambda) $, where $ \uS $ and $ \uLambda $ are $ \NR \times \NR $ matrices. Then, new expressions for $ {{_0\!F_0}} (\uS, \uLambda) $ are derived in terms of the determinant of a matrix with elementary-function entries.
Specializing to the case of Rician$(1)$/Rayleigh$(\NT-1)$ fading yields a new determinantal expression for the Stream-1 SNR m.g.f..
Comparing the old and new SNR m.g.f.~expressions reveals, for when $ \uS $ has single nonzero eigenvalue $ \sigma_1 $ and $ \uLambda $ is rank-$ N $ idempotent, the previously-unknown hypergeometric function relationship $ {{_0\!F_0}} (\uS, \uLambda) = {_1F_1} \left(N; \NR; \sigma_1 \right) $, and, consequently, a new determinantal expression for the latter.

Numerical results obtained from AEP expressions deduced from the newly-derived exact SNR-m.g.f.~expressions reveal the intriguing effect of the mean--correlation \textsl{condition} on the relative performance of MIMO ZF for Rician fading vs.~Rayleigh-only fading, and of the interfering-fading type and fading correlation on intended-stream performance.

\subsection{Notation}
\label{section_notation}

Scalars, vectors, and matrices are represented with lowercase italics, lowercase boldface, and uppercase boldface, respectively, e.g., $ h $, $ \uh $, and $ \uH $; 
$ \uh \sim {\cal{CN}} (\uh_{\text{d}}, \uR $) indicates that $ \uh $ is a complex-valued circularly-symmetric Gaussian random vector\cite[p.~39]{paulraj_book_05}\cite{gallager_08_prep} with mean (i.e., deterministic component) $ \uh_{\text{d}} $ and covariance matrix $ \mbR $; 
$ \uH \doteq \NR \times \NT $ indicates that matrix $ \uH $ has $ \NR $ rows and $ \NT $ columns;
$ \uH \sim {\cal{CN}} \left( \uHd, \mbI_{\NR} \otimes \uRTK \right) $ indicates that $ \uH $ is complex circularly-symmetric Gaussian matrix with mean $ \uHd $ and transmit-side covariance matrix $ \uRTK $\cite{mckay_tit_05};
$ r = \text{rank} (\uHd) $; 
subscripts $ \cdot_{\text{d}} $ and $ \cdot_{\text{r}} $ identify deterministic and random components, respectively; subscript $ \cdot_{\text{norm}} $ indicates a normalized variable; 
$ 1 : N $ stands for the enumeration $ 1, \, 2, \, \cdots , \, N$; 
superscripts $ \cdot^{\cal{T}} $ and $ \cdot^{\cal{H}} $ stand for transpose and Hermitian (i.e., complex-conjugate) transpose, respectively;
$ [\uH]_{i,j} $ indicates the $i,j$th (scalar) element of matrix $ \uH $; 
$ {\cal{CW}}_{\NT} \left(\NR, \uRT \right) $ denotes the complex CWD with dimension $ \NT $, degrees of freedom $ \NR $, and scale matrix $ \uRT $;
$ {\cal{CW}}_{\NT} \left(\NR, \uRTK, \uRTKinv \uHdH \uHd \right) $ denotes the complex NCWD with dimension $ \NT $, degrees of freedom $ \NR $, scale matrix $ \uRTK $, and noncentrality parameter matrix $ \uRTKinv \uHdH \uHd $\cite{mckay_tit_05};
$ {\cal{G}} (N, \Gamma) $ denotes the Gamma distribution with shape $ N $ and scale $ \Gamma $;
$ \uHa $ and $ \uHb $ are the submatrices obtained by partitioning $ \uH $ along its columns as in $ \uH = ( \begin{matrix} \uHa & \uHb \end{matrix} )$; 
accordingly, $ \uWaa $, $ \uWab $, $ \uWba $, and $ \uWbb $ are the submatrices obtained by partitioning the Gramian matrix $ \uW = \uHH \uH $ along rows and columns, so that $ \uW_{i,j} = \uH_i^{\cal{H}} \uH_j $, with $i, j = 1, 2 $;
$ \uWinvaa $, $ \uWinvab $, $ \uWinvba $, and $ \uWinvbb $ are the submatrices obtained by partitioning $ \uWinv $; 
the SC of $ \uW_{22} $ in $ \uW $ is given by $ \uGamma_1 = (\uW^{11})^{-1} = \uW_{11} - \uW_{12} \uW_{22}^{-1} \uW_{21} $\cite{zhang_book_05}\cite{ouellette1_laa_81};
$ \uGamma_1 | \uHb $ represents the random matrix $ \uGamma_1 $ conditioned on matrix $ \uHb $;
$ \| \uH \|^2 = \sum_{i}^{\NR} \sum_{j}^{\NT} | [\uH]_{i,j} |^2 = \text{tr}(\uH^{\cal{H}} \uH) $ is the squared Frobenius norm of $ \uH $; 
$ \text{tr} (\uX) $ represents the trace of $ \uX $, and $\text{etr} ( \uX) = e^{\text{tr}{(\uX)}}$; 
$ \mzero $ is the zero vector or matrix of appropriate dimensions; 
$ \text{diag} (\cdot, \, \cdots, \cdot ) $ is the diagonal matrix with given elements;
$ \mathbb{E} \{ \cdot \} $ denotes statistical average; 
$ \,{\buildrel d \over =}\, $ and $ \,{\buildrel d \over \approx}\, $ relate random variables with the same and approximately the same distribution, respectively;
$ {{_0\!F_0}} (\uS) $ is the hypergeometric function with a single matrix argument defined in\cite[Eq.~(35.8.1), p.~772]{NIST_book_10} and characterized by $ {{_0\!F_0}} (\uS) = \text{etr} ( \uS ) $\cite[Eq.~(35.8.2)]{NIST_book_10};
$ {{_0\!F_0}} (\uS, \uLambda) $ is the hypergeometric function of double matrix argument defined in\cite[Eq.~(88)]{james_ams_64}\cite[Eq.~(9)]{chiani_tit_10};
$ {_1\!F_1} (\cdot; \cdot; \sigma_1) $ is the confluent hypergeometric function of scalar argument $ \sigma_1 $\cite[Eq.~(13.2.2), p.~322]{NIST_book_10}; 
$ (N)_n $ is the Pochhammer symbol, i.e.,  $ (N)_0 = 1 $ and  $ (N)_n = N (N + 1) \ldots (N + n - 1) $, $ \forall n > 1 $\cite[p.~xiv]{NIST_book_10};
finally, $ \Rightarrow $ and $ \Leftrightarrow $ represent implication and equivalence, respectively, whereas `iff' is short for `if and only if'.

\subsection{Paper Outline}
Section~\ref{section_system_channel_model} shows our model and details our assumptions. 
Section~\ref{section_Schur_Complement} explains the SC--SNR relationship and characterizes the conditioned SC as NCWD.
Section~\ref{section_Rician_Rician_Condition} reveals the mean--correlation \textit{condition} for the SC to be CWD and for ZF SNRs to be Gamma-distributed.
Section~\ref{section_approximation} shows that the obtained Gamma distribution coincides with the virtual Gamma distribution under the revealed  \textit{condition}.
Section~\ref{section_uGamma1_distribution_Rician_Rayleigh_v_1} characterizes the matrix-SC distribution for Rician$(v)$/Rayleigh$(\NT-v)$ fading.
Finally, Section~\ref{section_Numerical_Results} presents numerical results.

\section{Signal, Channel, and Noise Models}
\label{section_system_channel_model}


Similarly to\cite{siriteanu_tvt_11}\cite{siriteanu_twc_13}, this paper considers an uncoded MIMO system over a frequency-flat fading channel.
There are $ \NT $ and $ \NR $ antenna elements at the transmitter(s) and receiver, respectively, with $ \NT \le \NR $.
Letting $\ux = [x_1 \, x_2 \, \cdots \, x_{\NT}]^{\cal{T}} \doteq \NT \times 1 $ denote the zero-mean transmit-symbol vector with $ \mathbb{E} \{ \ux \ux^{\cal{H}} \} = \mbI_{\NT} $, the vector with the received signals is\cite[p.~63]{paulraj_book_05}
\begin{eqnarray}
\label{equation system}
\ur = \sqrt{\frac{E_{\text{s}}}{\NT}} \, \uH \ux + \un \doteq \NR \times 1.
\end{eqnarray}
Above, $ E_{\text{s}}/\NT $ is the energy transmitted per symbol (i.e., per antenna), so that $ E_{\text{s}} $ is the energy transmitted per channel use.
The additive noise vector $\un$ is uncorrelated, circularly-symmetric, zero-mean, complex Gaussian with $\un \sim {\cal{CN}} (\mzero, N_0 \, \mbI_{\NR})$\cite[p.~39]{paulraj_book_05}\cite{gallager_08_prep}; $ \untilde = \un/\sqrt{N_0} \sim {\cal{CN}} (\mzero, \mbI_{\NR} ) $ will also be employed.
Then, the per-symbol transmit-SNR is
\begin{eqnarray}
\label{equation_rho}
\Gamma_{\text{s}} = \frac{E_{\text{s}}}{N_0} \frac{1}{\NT}.
\end{eqnarray}
In~(\ref{equation system}), $\uH \doteq \NR \times \NT $ is the complex-Gaussian channel matrix, assumed to have rank $ \NT $.
Its deterministic and random components are denoted as $\uHd$ and $\uHr$, respectively.
The channel matrix for Rician fading is usually written as\cite[p.~41]{paulraj_book_05}
\begin{eqnarray}
\label{equation channelH}
\uH =  \uHd + \uHr = \sqrt{\frac{K}{K+1}} \, \uH_{\text{d,norm}} + \sqrt{\frac{1}{K+1}} \, \uH_{\text{r,norm}},
\end{eqnarray}
where it is assumed, for normalization purposes\cite{loyka_twc_09}, that:
\begin{eqnarray}
\label{equation_normalization}
\| \uH_{\text{d,norm}} \|^2 = \mathbb{E} \{ \| \uH_{\text{r,norm}}  \|^2 \} = \NR \NT.
\end{eqnarray}
Then, if $ [\uHd]_{i,j} = 0 $, $ |\left[ \uH \right]_{i,j}| $ is Rayleigh-distributed; otherwise, $ |\left[ \uH \right]_{i,j}| $ is Rician-distributed\cite{simon_alouini_book_05}, and the power ratio
\begin{eqnarray}
\label{equation K_definition}
K = \frac{\| \uHd \|^2}{\mathbb{E} \{ \| \uHr \|^2 \} } = \frac{ \frac{K}{K+1} \| \uH_{\text{d,norm}} \|^2}{\frac{1}{K+1} \mathbb{E} \{ \| \uH_{\text{r,norm}} \|^2 \} }
\end{eqnarray}
is the Rician $ K $-factor.

For analysis tractability, we assume, as in \cite{kiessling_spawc_03}\cite{gore_cl_02}, that the receive-side correlation is zero and that any row $ \ug^{\cal{H}}_{\text{r,norm}} $ of $ \uH_{\text{r,norm}} $ is distributed as $ \ug_{\text{r,norm}}  \sim {\cal{CN}} (\mzero, \uRT ) $, where $ \uRT $ is Hermitian (i.e., $ \uRT = \uRTH $).
Considering independent $ [\uH_{\text{r,w,norm}}]_{i, j} \sim {\cal{CN}} \left( 0, 1 \right)  $, $ i = 1 : \NR, j = 1 : \NT $, we can write $ \uH_{\text{r,norm}} = \uH_{\text{r,w,norm}} \uRTsqrt $, which helps show that $  \mathbb{E} \{ \| \uH_{\text{r,norm}}  \|^2 \} = \NR \NT \Leftrightarrow \trace(\uRT) = \NT $.
Therefore, our normalization model~(\ref{equation_normalization}) allows for unequal $ [\uRT]_{i,i} $, $ i = 1 : \NT $, as long as $ \sum_{i = 1}^{\NT} [\uRT]_{i,i}  = \NT $.
Thus, the model allows for $ \mathbb{E} \{ | [\uH_{\text{r,norm}}]_{i,j}  |^2 \} \neq  \mathbb{E} \{ | [\uH_{\text{r,norm}}]_{i,k}  |^2 \} $, $ \forall i = 1 : \NR $, $ \forall j \neq k $. 
On the other hand, the elements of $ \uH_{\text{d,norm}} $ may have different amplitudes and phases as long as the entire matrix satisfies $ \| \uH_{\text{d,norm}} \|^2 = \NR \NT $. 
In conclusion, our analysis applies for both collocated and non-collocated transmit antennas.\label{pageref_collocated}

Based on the above assumptions, any row $ \ug_{\text{r}}^{\cal{H}} $ of $ \uHr $ is characterized by $ \ug_{\text{r}} \sim {\cal{CN}} (\mzero, \uRTK ) $, where\cite[Eq.~(5)]{siriteanu_tvt_11}
\begin{eqnarray}
\label{equation_RTK_Hr}
\uRTK = \mathbb{E} \{ \ug_{\text{r}} \ug_{\text{r}}^{\cal{H}} \} = \frac{1}{\NR}\mathbb{E} \{ \uHrH \uHr\} = \frac{1}{K + 1 } \uRT,
\end{eqnarray}
so that $ \uH \sim {\cal{CN}} \left( \uHd, \mbI_{\NR} \otimes \uRTK \right) $\cite{mckay_tit_05}.

Matrix $ \uRT $ can be computed from the azimuth spread (AS) as shown in\cite[Section~VI.A]{siriteanu_tvt_11}, when assuming Laplacian power azimuth spectrum, as adopted in WINNER II\cite{winner_d_1_1_2_v_1_2}.
Measured AS (in degrees) and $ K $ were modeled in WINNER II with scenario-dependent lognormal distributions.


\section{ZF SNR Relationship with Schur Complement in Wishart Gramian Matrix}
\label{section_Schur_Complement}

\subsection{Matrix Partitionings and Related Equalities}
\label{section_partitioning}

We introduce below a series of matrix partitionings, decompositions, and ensuing relationships that will be employed throughout.
In\cite{siriteanu_twc_13} we employed the vector--matrix partition
\begin{eqnarray}
\label{equation_partitioned_H_general_v_1}
\uH = \begin{matrix} ( {\uh_1} & {\uH_2})  \end{matrix} = \begin{matrix} ( \uhda \quad \uHdb \end{matrix} ) + \begin{matrix} ( \uhra \quad \uHrb \end{matrix} ),
\end{eqnarray}
where $ \uh_1 $, $ \uhda $, and $ \uhra $ are $ \NR \times 1 $ vectors, whereas $ \uH_2 $, $ \uHdb $, $ \uHrb $ are $ \NR \times (\NT - 1) $ matrices.
However, partitioning~(\ref{equation_partitioned_H_general_v_1}) can help characterize only the performance for the transmitted stream affected by vector $ \uh_1 $, referred to herein as Stream 1. 

Herein, we employ instead the matrix--matrix partitioning
\begin{eqnarray}
\label{equation_partitioned_H_general}
\uH = \begin{matrix} ( {\uH_1} & {\uH_2} ) \end{matrix} = \begin{matrix} ( \uHda \quad \uHdb \end{matrix} ) + \begin{matrix} ( \uHra \quad \uHrb \end{matrix} ),
\end{eqnarray}
where $ \uH_1 $, $ \uHda $, and $ \uHra $ are $ \NR \times v $ matrices, whereas $ \uH_2 $, $ \uHdb $, $ \uHrb $ are $ \NR \times (\NT - v) $ matrices, with $ 1 \le v < \NT $. 
According to~(\ref{equation_partitioned_H_general}), we partition the column-sample-correlation matrix of $ \uH $, i.e., the Gramian $ \uW  = \uHH \uH $, and its inverse $ \uWinv $ as mentioned in Section~\ref{section_notation}.
We also partition the covariance matrix $ \uRTK $ into its component submatrices $ \RTaa $, $ \RTab $, $ \RTba $, and $ \RTbb $, where $ \RTba = \RTabH  $.
Also, we partition $ \uRTKinv $ into its component submatrices $ \RTupaa $, $ \RTupab $, $ \RTupba $, and $ \RTupbb $.
Further, for $ \uRTK $ we consider the upper--lower triangular (UL) decomposition $ \uRTK = \uA \uAH $\cite[Sec.~5.6]{gentle_book_07}, and partition the upper triangular matrix $ \uA $ into its component submatrices $ \uAaa $, $ \uAab $, $ \uAba = \mzero $, and $ \uAbb $
Finally, we partition $ \uAinv $ into its component submatrices $ \uAinvaa $, $ \uAinvab $, $ \uAinvba  = \mzero $, and $ \uAinvbb $. 
For these matrices we have deduced the following relationships, for subsequent use:
\begin{eqnarray}
\label{equation_A21A11A22A21_our}
\uA_{11}^{-1} & = & \uA^{11}, \; \uA_{22}^{-1} = \uA^{22}, \uA^{12} = - \uA^{11} \uA_{12} \uA^{22} \\
\RTbbinva & = & (\uA_{22} \uA_{22}^{\cal{H}})^{-1} = \uA_{22}^{-\cal{H}} \uA_{22}^{-1} = \uA^{22,{\cal{H}}} \uA^{22} \\ 
\label{equation_RTK11inv_our}
\RTba & = & \uA_{22} \uA_{12}^{\cal{H}} \\
\label{equation_A11_A11H}
\uA_{11} \uA_{11}^{\cal{H}} & = & (\uA^{11,\cal{H}} \uA^{11})^{-1} = 
(\RTupaa)^{-1} \\
\label{equation_Schur_Complement_RT} & = & \RTaa - \RTab \, \RTbbinva \, \RTba.
\end{eqnarray}

\begin{remark}
\label{remark_Schur_complement}
The matrix described by~(\ref{equation_A11_A11H}) and (\ref{equation_Schur_Complement_RT}) is referred to as the Schur complement (SC) of $ \RTbb $ in $ \uRTK $\cite{zhang_book_05}\cite{ouellette1_laa_81}\cite[Sec.~3.4]{gentle_book_07}\cite[Appendix]{kiessling_spawc_03}. 
For our channel model, it represents the correlation of the first $ v $ elements of $ \ug_{\text{r}} $ given its remaining $ \NT - v $ elements.
\end{remark}

%

\subsection{Schur Complement in the Gramian Matrix $ \uW $}
\label{section_Wishart_Schur}

The SC of $ \uW_{22} = \uHbH \uHb $ in Gramian $ \uW $ is the matrix
\begin{eqnarray}
\label{equation_Schur_Complement_W}
\uGamma_1 = (\uW^{11})^{-1} = \uW_{11} - \uW_{12} \uW_{22}^{-1} \uW_{21} \doteq v \times v.
 \end{eqnarray}
It can be expressed as a matrix Hermitian form as follows:
\begin{eqnarray}
\label{equation_Schur_Complement_W_expanded}
\uGamma_1  & = & \uHaH \uHa  - \uHaH \uHb (\uHbH \uHb)^{-1} \uHbH \uHa \\
\label{equation_Q_definition}
& = & \uHaH \underbrace{\left[  \mbI_{\NR} - \uHb \left( \uHbH \uHb \right)^{-1} \uHbH  \right]}_{=\uQ_2} \uHa.
 \end{eqnarray}
First, note from~(\ref{equation_Schur_Complement_W_expanded}) that the SC matrix $ \uGamma_1  $ is the column sample-correlation of $ \uHa $ given  $ \uHb $. 
Then, note that matrix $ \uHb \left( \uHbH \uHb \right)^{-1} \uHbH \doteq \NR \times \NR $ is the projection onto the column space of $ \uHb $, whereas  matrix $ \uQ_2 \doteq \NR \times \NR $ is the projection onto the null space of $ \uHbH $.
These Hermitian matrices are idempotent and have eigenvalues as listed below:
\begin{eqnarray}
\label{equation_Projection_H2}
\uHb \left( \uHbH \uHb \right)^{-1} \uHbH: & 1, \; 1, \; \cdots, \; 1, & 0, \; 0, \; \cdots, \; 0\\
\label{equation_Projection_Q2}
\uQ_2: & \underbrace{0, \; 0, \; \cdots, \; 0,}_{\NT-v} & \underbrace{1, \; 1, \; \cdots, \; 1}_{\NR - \NT + v = N_v}
\end{eqnarray}
Their ranks are $ \NT - v $ and $ N_v $, respectively.
We shall denote $ N_v $ for $ v = 1$, i.e., $ \NR - \NT + 1 $, simply as $ N $, as in\cite{siriteanu_twc_13}.

\subsection{Relationship of $ \uGamma_1 $ with ZF SNRs}
\label{section_SNR_intro}
Given $\uH$ and nonsingular $ \uW = \uH^{\cal{H}} \uH $, ZF for the signal from~(\ref{equation system}) means separately mapping into the closest modulation (e.g., $M$PSK) constellation symbol each element of the following $ \NT \times 1 $ vector\cite[p.~153]{paulraj_book_05}:
\begin{eqnarray}
\label{equation ZF_for_perfect_CSI}
\uy = \sqrt{\frac{ \NT }{ E_{\text{s}} }} \left[ \uH^{\cal{H}} \uH \right]^{-1} \uH^{\cal{H}} \, \ur = \ux + \frac{ 1 }{ \sqrt{\Gamma_{\text{s}} }} \left[ \uH^{\cal{H}} \uH \right]^{-1} \uH^{\cal{H}} \untilde.
\end{eqnarray}
Since the resulting noise vector has correlation matrix $ \uW^{-1}/{ \Gamma_{\text{s}}} $,
the ZF SNR for Stream $ i = 1 : \NT $ has usually been expressed in ratio form as follows\cite[p.~153]{paulraj_book_05}\cite{gore_cl_02}
\begin{eqnarray}
\label{equation_gammak_perfect_CSI}
\gamma_i = \frac{ \Gamma_{\text{s}} }{\left[\uW^{-1}\right]_{i,i}}.
\end{eqnarray}
Now, $ \forall v = 1 : \NT $ we can write, based on~(\ref{equation_Schur_Complement_W}), that
\begin{eqnarray}
\label{equation_gamma_from_W_G}
\gamma_i = \frac{ \Gamma_{\text{s}} }{\left[\uW^{-1}\right]_{i,i}} = \frac{\Gamma_{\text{s}}}{[\uW^{11} ]_{i,i}} =  \frac{\Gamma_{\text{s}}}{[\uGamma_1^{-1}]_{i,i}}, \quad i = 1 : v.
\end{eqnarray}
Thus, in general ($ \forall v $), the ZF SNRs for Streams $ i = 1 : v $ are determined by the SC matrix $ \uGamma_1 $, through its inverse $ \uGamma_1^{-1} $.
Only for $ v = 1 $, i.e., when $ \uGamma_1 $ reduces to a scalar, we can write the ZF SNR for Stream 1 in terms of the SC as\cite{siriteanu_twc_13}
\begin{eqnarray}
\label{equation_gamma1_Gamma1}
\gamma_1 = \frac{ \Gamma_{\text{s}} }{\left[\uW^{-1}\right]_{1,1}} = \frac{\Gamma_{\text{s}}}{\uW^{11}} = {\Gamma_{\text{s}}}{(\uW^{11})^{-1}} = \Gamma_{\text{s}} \uGamma_1.
\end{eqnarray}
Based on~(\ref{equation_Schur_Complement_W_expanded}), we can put the SC in scalar Hermitian form
\begin{eqnarray}
\label{equation_gamma1_Gamma1_HF}
\uGamma_1 = \uhaH \uQ_2 \uha,
\end{eqnarray}
which has helped characterize the distribution of $ \gamma_1 $ for Rician$ (1) $/Rayleigh$(\NT - 1)$ fading as an infinite linear combination of Gamma distributions in\cite[Eq.~(37)]{siriteanu_twc_13} --- see also~(\ref{equation_gamma1_mgf_final}) and~(\ref{equation_1F1_definition}), on page~\pageref{equation_gamma1_mgf_final}.

For the more general case $ v \ge 1 $, we analyze hereafter in this paper the distribution of the matrix-SC $ \uGamma_1 $ based on its Hermitian form~(\ref{equation_Q_definition}) and exploit its relationship from~(\ref{equation_gamma_from_W_G}) with ZF SNRs to analyze the ZF performance for Streams $ i = 1 : v $. 

\subsection{Distribution of $ \uGamma_1 $ Conditioned on $ \uHb $ (or $ \uQ_2 $)}
\label{section_Gamma1_cond_distrib}
Nonzero- and zero-mean complex-Gaussian $ \uH $ yield complex NCWD and CWD Gramian $ \uW $, respectively\cite{mckay_tit_05}:
\begin{eqnarray}
\label{equation_noncentral_Wishart_distribution}
&& \uH \sim {\cal{CN}} \left( \uHd, \mbI_{\NR} \otimes \uRTK \right) \Rightarrow\uW \sim {\cal{CW}}_{\NT} \left(\NR, \uRTK, \uRTKinv \uHdH \uHd \right) \\
\label{equation_central_Wishart_distribution}
&& \uH \sim {\cal{CN}} \left( \mzero, \mbI_{\NR} \otimes \uRT \right) \Rightarrow\uW \sim {\cal{CW}}_{\NT} \left(\NR, \uRT \right).
\end{eqnarray}
Because $ \uHa $ and $ \uHb $ are jointly Gaussian, the distribution of $ \uHa $ given $ \uHb $ is\cite[Appendix]{kiessling_spawc_03}
\begin{eqnarray}
\label{equation_distribution_H1_given_H2}
\uHa | \uHb \sim
{\cal{CN}} \big( \uM + \uHb \uR_{2,1}, \mbI_{\NR} \otimes \big( \RTupaa \big)^{-1} \big),
\end{eqnarray}
\footnote{This corroborates Remark~\ref{remark_Schur_complement} on the meaning of SC $ \big( \RTupaa \big)^{-1} $. }where 
\begin{eqnarray}
\label{equation_M}
\uM & = & \uHda - \uHdb \uR_{2,1} \doteq \NR \times v,\\
\label{equation_R21}
\uR_{2,1} & = & \RTbbinva \RTba \doteq (\NT - v) \times v,
\end{eqnarray}
are deterministic matrices. We can now recast~(\ref{equation_distribution_H1_given_H2}) further as
\begin{eqnarray}
\label{equation_ha_Htilde}
\uHa | \uHb \,{\buildrel d \over =}\,  \uX + \uHb \uR_{2,1}; \uX \sim {\cal{CN}} \left( \uM, \mbI_{\NR} \otimes \left( \RTupaa \right)^{-1} \right).
\end{eqnarray}
Substituting this in~(\ref{equation_Q_definition}) and manipulating as in\cite{kiessling_spawc_03} yields
\begin{eqnarray}
\uGamma_1 | \uQ_2 \,{\buildrel d \over =}\,  \uXH \uQ_2 \uX.
\end{eqnarray}
This matrix Hermitian form has, for $ \uM \neq \mzero $, the NCWD\cite[Cor.~7.8.1.1, p.~255]{gupta_book_00}
\begin{eqnarray}
\label{equation_gamma1_cond_q}
\uGamma_1 | \uQ_2 \sim {\cal{CW}}_{v} \left(N_v, \left( \RTupaa \right)^{-1}, \RTupaa \uMH \uQ_2 \uM \right).
\end{eqnarray}
Thus, its m.g.f.~for matrix $ \uTheta \doteq v \times v $ is given by\cite[Eq.~(4)]{kuriki_aism_10}
\begin{eqnarray}
\label{equation_uGamma1_mgf}
M_{\uGamma_1|\uQ_2}(\uTheta)
= {\big |  \mbI_{v} - \uTheta \left( \RTupaa \right)^{-1} \big |^{-N_v}} { \text{etr} \bigg( \left[ \mbI_{v} - \uTheta \left( \RTupaa \right)^{-1}   \right]^{-1} \uTheta \uMH \uQ_2 \uM  \bigg ) }.
\end{eqnarray}
Deriving from~(\ref{equation_uGamma1_mgf}) the m.g.f.~of the unconditioned $ \uGamma_1 $ as
\begin{eqnarray}
\label{equation_uGamma1_mgf_avg}
M_{\uGamma_1}(\uTheta) 
= {\big |  \mbI_{v} - \uTheta \left( \RTupaa \right)^{-1} \big |^{-N_v}} \mathbb{E}_{\uQ_2} \bigg\{  { \text{etr} \bigg( \left[ \mbI_{v} - \uTheta \left( \RTupaa \right)^{-1}   \right]^{-1} \uTheta \uMH \uQ_2 \uM  \bigg ) } \bigg\}
\end{eqnarray}
remains intractable for general Rician fading.
Nevertheless,~(\ref{equation_uGamma1_mgf_avg}) yields the distribution of the unconditioned $ \uGamma_1 $ for the following special cases:
\begin{enumerate}
\item Full-Rician fading under condition $ \uM = \mzero $. Note that $ \uM = \mzero $ covers the trivial case of Rayleigh-only fading as well as a case that may be practically relevant: Rayleigh$ (v) $/ Rician$(\NT - v)$ fading whereby the Rayleigh fading is uncorrelated with the Rician fading.
\item Rician$ (v) $/Rayleigh$(\NT - v)$ fading.
\end{enumerate}
They are treated in Sections~\ref{section_Rician_Rician_Condition} and~\ref{section_uGamma1_distribution_Rician_Rayleigh_v_1}, respectively. 


\section{Distribution of ZF SNRs  for Streams $ i = 1 : v $, $ v = 1 : \NT $, Under Rician Fading with $ \uM = \mzero $}
\label{section_Rician_Rician_Condition}

\subsection{$ \uGamma_1 $ is CWD If and Only If $ \uM = \mzero $, i.e., $ \uHda = \uHdb \uR_{2,1} $}
\label{section_Rician_Rician_Condition_Equivalence}

The theorem below follows readily from the fact that in~(\ref{equation_uGamma1_mgf_avg}) $ \mathbb{E}_{\uQ_2} \{ \text{etr}(\cdot) \} = 1 $  \textit{iff} $ \uM =  \uHda - \uHdb \uR_{2,1} = \mzero $.


\begin{theorem}
\label{theorem_Gamma1_central_condition}
\begin{eqnarray}
\label{equation_special_condition_Hd1_Hd2_R21}
&& \uHda = \uHdb \uR_{2,1} \\
& \Leftrightarrow & M_{\uGamma_1}(\uTheta) = {\big | \mbI_{v} - \uTheta \left( \RTupaa \right)^{-1}  \big |^{-N_v}} \\
\label{equation_uGamma1_distribution_M_zero}
& \Leftrightarrow &  \uGamma_1 \sim  {\cal{CW}}_{v} \left(N_v, \left( \RTupaa \right)^{-1} \right).
\end{eqnarray} 
\end{theorem}

\begin{remark}
\label{remark_cond_Ray_Rice_1}
The mean--correlation condition~(\ref{equation_special_condition_Hd1_Hd2_R21}) holds for: 
\begin{itemize}
\item Rayleigh-only fading, i.e., $ \uHda = \mzero $ and $ \uHdb = \mzero $ (then, the value of $ v $ is irrelevant). 
\item Rayleigh$(v)$/Rician$ (\NT-v) $ fading, i.e., for $ \uHda = \mzero $, $ \uHdb \neq \mzero $, if the Rayleigh fading is uncorrelated with the Rician fading, which reduces to zero $ \RTba $, i.e., $ \uR_{2,1} $, in~(\ref{equation_R21}). 
\end{itemize}
\end{remark}
For full-Rician fading, condition~(\ref{equation_special_condition_Hd1_Hd2_R21}) implies an interesting ``parallelism" between means and correlations, as shown in Appendix~\ref{section_further_corollaries}, Corollary~\ref{corollary_parallelism}. That Appendix provides some additional analysis and insights into condition~(\ref{equation_special_condition_Hd1_Hd2_R21}).


\subsection{ZF SNR Distribution for Streams $ i = 1 : v $}
\label{section_ZF_SNR_Distribution}

For CWD $ \uGamma_1 $, i.e., for $ \uHda = \uHdb \uR_{2,1} $, the following Lemma characterizes as Gamma-distributed the ZF SNRs for Streams $ i = 1 : v $.

\begin{lemma}
\label{lemma_Gamma1_central_gamma_i_Gamma}
\begin{eqnarray}
\label{equation_lemma2}
&& \uGamma_1 \sim  {\cal{CW}}_{v} \left(N_v, \left( \RTupaa \right)^{-1} \right) \Rightarrow \text{for } i = 1 : v \nonumber \\ 
\label{equation_gamma_distribution_Gamma_exact}
&& \gamma_i = \frac{\Gamma_{\text{s}}}{[\uGamma_1^{-1}]_{i,i}} \sim {\cal{G}} \left(N, \Gamma_{K,i} = \frac{\Gamma_{\text{s}}}{\left[ \uRTKinv \right]_{i,i}} \right). \quad \quad
\end{eqnarray}
\end{lemma}

\IEEEproof{A special case of\cite[Th.~3.2.11, p.~95]{muirhead_book_05} yields
\begin{eqnarray}
\label{equation_lemma2_proof}
&& \uGamma_1 \sim  {\cal{CW}}_{v} \left(N_v, \left( \RTupaa \right)^{-1} \right) \Rightarrow  \text{for } i = 1 : v \nonumber \\ 
&& \frac{1}{[\uGamma_1^{-1}]_{i,i}} \sim {\cal{CW}}_{1} \bigg(N, \frac{1}{\left[ \RTupaa \right]_{i,i}} \bigg). \nonumber
\end{eqnarray}
Because for $ i = 1 : v $ we can write $ \left[\RTupaa \right]_{i,i} = \left[ \uRTKinv \right]_{i,i} $, we can express the m.g.f.~of $ {1}/{[\uGamma_1^{-1}]_{i,i}} $ as\cite[Eq.~(4)]{kuriki_aism_10}
\begin{eqnarray}
\label{equation_lemma2_proof_chi_squared}
M(s) = \left( 1 - s /\left[ \uRTKinv \right]_{i,i} \right)^{-N},
\end{eqnarray}
which yields the desired result, i.e.,
\begin{eqnarray}
\label{equation_gamma_mgf_Gamma_exact}
M_{\gamma_i}(s) = M(s \Gamma_{\text{s}}) = \left( 1 - s \Gamma_{K,i} \right)^{-N}.
\end{eqnarray}
}

\begin{corollary}
\label{corollary_Ray_Rice}
For Rayleigh-only fading ($ K = 0 $), the ZF SNRs on all streams $ i = 1 : \NT $ are Gamma-distributed as:
\begin{eqnarray}
\label{equation_gamma_distribution_Gamma_exact_Rayleigh}
&& \gamma_i = \frac{\Gamma_{\text{s}}}{[\uGamma_1^{-1}]_{i,i}} \sim {\cal{G}}  \left( N, \Gamma_{0,i} = \frac{\Gamma_{\text{s}}}{\left[ \RT^{-1} \right]_{i,i}} \right), \quad \quad
\end{eqnarray}
whereas for some Rician fading ($ K \neq 0 $) satisfying $ \uM = \mzero $, Streams $ i = 1 : v $ are Gamma-distributed as in~(\ref{equation_gamma_distribution_Gamma_exact}), with\footnote{Based on~(\ref{equation_RTK_Hr}).}
\begin{eqnarray}
\label{equation_GammaKi}
\Gamma_{K,i} = \frac{\Gamma_{\text{s}}}{\left[ \uRTKinv \right]_{i,i}} = \frac{1}{K+1}  \frac{\Gamma_{\text{s}}}{\left[ \RT^{-1} \right]_{i,i}} = \frac{1}{K+1}  \Gamma_{0,i}.
\end{eqnarray}
\end{corollary}

\begin{remark}
\label{remark_M_zero}
If $ \uM = \mzero $ then Rician fading on any stream
\begin{itemize}
\item does not change SNR distribution type (Gamma) for Streams $ i = 1 : v $, compared to Rayleigh-only fading; these SNR distributions are also independent of $ v $, $  \uHd $.
\item reduces the average SNR ($ \mathbb{E} \{ \gamma_i \} = N \Gamma_{K,i} $) for Streams $ i = 1 : v $ by a factor of $ K + 1$ over Rayleigh-only fading; this is illustrated numerically in Section~\ref{section_Numerical_Results}.
\item leaves intractable the derivation of the ZF SNR distributions for streams $ i = v + 1 : \NT $.
\end{itemize}
\end{remark}

\subsection{AEP Expression for Streams $ i = 1 : v $}
\label{section_AEP_Expression}

Knowing the SNR m.g.f., the elegant AEP-derivation procedure from\cite[Ch.~9]{simon_alouini_book_05} can be employed, e.g., for $ M $PSK modulation.
Because then the Stream-$ i $ error probability is given by\cite[Eq.~(8.22)]{simon_alouini_book_05}
\begin{eqnarray}\label{equation_instantaneous_Pe}
P_{\text{e}}(\gamma_i) = \frac{1}{\pi} \int_{0}^{\frac{M-1}{M}\pi}
\exp\left \{{-\gamma_i \, \frac{ \sin^2{\frac{\pi}{M}}}{\sin^2\theta}} \right \} \nid \theta,
\end{eqnarray}
the AEP can be written as\cite[Chapter~9]{simon_alouini_book_05}
\begin{eqnarray}
\label{equation_average_Pe}
P_{\text{e},i} = \mathbb{E} \{ P_{\text{e}}(\gamma_i) \}  = \frac{1}{\pi} \int_{0}^{\frac{M-1}{M}\pi}
M_{\gamma_i}\left(-\frac{ \sin^2{\frac{\pi}{M}}}{\sin^2 \theta}\right) \nid \theta.
\end{eqnarray}
Substituting the m.g.f.~from~(\ref{equation_gamma_mgf_Gamma_exact}) into~(\ref{equation_average_Pe}) yields the \textsl{exact AEP expression} for Streams $ i = 1 : v $, under condition~(\ref{equation_special_condition_Hd1_Hd2_R21}),
\begin{eqnarray}
\label{equation_average_Pe_condition}
P_{\text{e},i}^{(\ref{equation_average_Pe_condition})} = \frac{1}{\pi} \int_{0}^{\frac{M-1}{M}\pi}
\left( 1 + \frac{ \sin^2{\frac{\pi}{M}}}{\sin^2 \theta} \Gamma_{K,i} \right)^{-N}\nid \theta.
\end{eqnarray}

\begin{center}
\begin{table*}[t]
{\small
\hfill{}
\caption{Dependence of ZF SNR Distribution and AEP on Fading Type and Mean--Correlation Condition $ \uHda = \uHdb \uR_{2,1} $ from~(\ref{equation_special_condition_Hd1_Hd2_R21})}
\renewcommand{\arraystretch}{1.35}
\label{table_gamma1_statistics}
\begin{tabular}{l|l|c|c|c|c|l|l}
  \hline
    & Fading Type & $ \uHda $ & $ \uHdb $ & $ \uR_{2,1} $ & (\ref{equation_special_condition_Hd1_Hd2_R21}) & $\gamma_i$ Distribution, $ i = 1 : v $ & AEP \\
   \hline \hline
   1 & Rayleigh-only & $ = \mzero $ & $ = \mzero $ & $ \forall $ & \cmark & $ \gamma_i \,{\buildrel d \over =}\, \widehat{\gamma}_i \sim {\cal{G}}  ( N, \Gamma_{0,i} ) $, see~(\ref{equation_gamma_distribution_Gamma_exact_Rayleigh}) & $ {P}_{\text{e},i}^{(\ref{equation_average_Pe_condition})} \! = \! \widehat{P}_{\text{e},i}^{(\ref{equation_average_Pe_approximation})}  $\\
  \hline 
   2 & Rayleigh$ (v) $/Rice$(\NT - v)$ & $ =\mzero $ & $ \neq \mzero $ & $  =\mzero $ & \cmark & $ \gamma_i \,{\buildrel d \over =}\, \widehat{\gamma}_i \sim {\cal{G}}  ( N, \Gamma_{K,i} ) $, see~(\ref{equation_lemma2}) & $ {P}_{\text{e},i}^{(\ref{equation_average_Pe_condition})} \! = \! \widehat{P}_{\text{e},i}^{(\ref{equation_average_Pe_approximation})}  $ \\
  \hline
    3 & Rice$ (v) $/Rice$(\NT - v)$ & $ \neq \mzero $ & $ \neq \mzero $ & $ \neq \mzero $ & \cmark & $ \gamma_i \,{\buildrel d \over =}\, \widehat{\gamma}_i \sim {\cal{G}}  ( N, \Gamma_{K,i} ) $, see~(\ref{equation_lemma2}) & $ {P}_{\text{e},i}^{(\ref{equation_average_Pe_condition})} \! = \! \widehat{P}_{\text{e},i}^{(\ref{equation_average_Pe_approximation})}  $ \\
  \hline \hline
    4 & Rice$ (1) $/Rayleigh$(\NT - 1)$ & $ \neq\mzero $ & $  =\mzero $ & $ \forall $ & \xmark & Known for $ \gamma_1 $, see~(\ref{equation_gamma1_mgf_final}),~(\ref{equation_gamma1_mgf_final_123}) & $ {P}_{\text{e},1}^{(\ref{equation_average_Pe_Rice_Ray_determinant})} $ \\
  \hline \hline
    5 & Rice$ (v) $/Rayleigh$(\NT - v)$, $ \!v\!>\!1$ & $ \neq\mzero $ & $  =\mzero $ & $ \forall $ & \xmark & Unknown; $ M_{\uGamma_1}(\uTheta) $ in~(\ref{equation_uGamma1_mgf_uU_1_avg_1}) & Unknown \\
  \hline
   6 & Rayleigh$ (v) $/Rice$(\NT - v)$ & $  =\mzero $ & $ \neq \mzero $ & $ \neq \mzero $ & \xmark &  Unknown  & Unknown \\
  \hline
    7 & Rice$ (v) $/Rice$(\NT - v)$ & $ \neq\mzero $ & $ \neq \mzero $ & $ \forall $ & \xmark & Unknown & Unknown \\
  \hline \hline
    8 & Virtual Rayleigh & $ \neq\mzero $ & $ \neq \mzero $ & $ \forall $ &   & $ \widehat{\gamma}_i  \sim {\cal{G}}  ( N, \widehat{\Gamma}_{K,i} ) $, see~(\ref{equation_gammahat_Gamma}) & $ \widehat{P}_{\text{e},i}^{(\ref{equation_average_Pe_approximation})}  $ \\
  \hline
\end{tabular}}
\hfill{}
\end{table*}
\end{center}

\subsection{Summary of Results}

In Table~\ref{table_gamma1_statistics}, Rows 1--3 characterize, based on Lemma~\ref{lemma_Gamma1_central_gamma_i_Gamma} and  Corollary~\ref{corollary_Ray_Rice}, ZF SNR distributions for fading cases whereby the mean--correlation condition~(\ref{equation_special_condition_Hd1_Hd2_R21}) holds (\cmark).

Remaining rows characterize fading cases whereby~(\ref{equation_special_condition_Hd1_Hd2_R21}) does not hold (\xmark). 
Of them, only for the case of Rician$ (1) $/Rayleigh$(\NT - 1)$ fading, characterized in Row 4, we have recently found in\cite{siriteanu_twc_13}, by partitioning with $ v = 1 $, that the exact distribution of $ \gamma_1 $ is an infinite linear combination of Gamma distributions\cite[Eq.~(37)]{siriteanu_twc_13} --- see also~(\ref{equation_gamma1_mgf_final}),~(\ref{equation_1F1_definition}).

In Section~\ref{section_uGamma1_distribution_Rician_Rayleigh_v_1}, we shall generalize the approach from\cite{siriteanu_twc_13} to the partitioning with $ v > 1 $, to express the m.g.f.~of $ \uGamma_1 $ under Rician$ (v) $/Rayleigh$(\NT - v)$ fading, which yields the determinantal expression for the m.g.f.~of $ \gamma_1 $ in~(\ref{equation_gamma1_mgf_final_123}), i.e., an alternative to the infinite-series expression\cite[Eq.~(37)]{siriteanu_twc_13}.

However, below, we first take a fresh look at a Wishart-distribution approximation\footnote{Characterized, for convenience, in Row 8 of Table~\ref{table_gamma1_statistics}.} proposed in\cite{steyn_roux_sasj_72} and applied for ZF analysis in\cite[Refs.~24-27,30,31]{siriteanu_tvt_11}, without accuracy testing.
Our numerical testing from\cite{siriteanu_tvt_11}  of this approximation revealed only that lower values of $ r = \text{rank} (\uHd) $ yield --- inconsistently --- lower ZF SNR distribution-approximation error.
Next, we reconsider this approximation analytically and reveal that it turns exact under condition~(\ref{equation_special_condition_Hd1_Hd2_R21}).

\section{Approximate and Exact Gamma Distributions for ZF SNRs}
\label{section_approximation}

\subsection{Approximate CWD for $ \uW $ Proposed in\cite{steyn_roux_sasj_72}}
\label{section_approximation_W}

On one hand, given the actual nonzero-mean channel matrix Rician fading $ \uH \sim {\cal{CN}} \left( \uHd, \mbI_{\NR} \otimes \uRTK \right) $, we have $ \uW = \uH^{\cal{H}} \uH \sim {\cal{CW}}_{\NT} \left(\NR, \uRTK, \uRTKinv \uHdH \uHd\right) $.
On the other hand, as in\cite{steyn_roux_sasj_72}, if we consider a virtual zero-mean matrix $ \uHhat \sim {\cal{CN}} \left( \mzero, \mbI_{\NR} \otimes \uRTKhat \right) $, then $ \widehat{\uW}  = \widehat{\uH}^{\cal{H}} \widehat{\uH} \sim {\cal{CW}}_{\NT} (\NR, \uRTKhat ) $.
The proof of the next Lemma follows from
\begin{eqnarray}
\label{equation_Wishart_same_mean}
\mathbb{E} \{ \uW \} =\NR \mbR_{\text{T},K} + \uHdH \uHd = \NR \uRTKhat = \mathbb{E} \{ \uWhat \}  .
\end{eqnarray}

\begin{lemma}
\label{theorem_Wishart_approximation}[{\cite{steyn_roux_sasj_72}\cite{siriteanu_tvt_11}}]
\begin{eqnarray}
\label{equation_Wishart_same_mean_1}
\label{equation_virtual_correlation_matrix}
\mathbb{E} \{ \uWhat \} = \mathbb{E} \{ \uW \} \Leftrightarrow
\widehat{\mbR}_{\text{T},K} = \mbR_{\text{T},K} + \frac{1}{\NR} \uHdH \uHd,
\end{eqnarray}
i.e., the two Wishart distributions have equal means iff relationship~(\ref{equation_virtual_correlation_matrix}) holds between the statistics of $ \uH $ and $ \uHhat $.
\end{lemma}


Based on the mean-equality~(\ref{equation_Wishart_same_mean_1}), the approximation of the NCWD of the actual $ \uW $ with the virtual CWD of $ \uWhat $
was proposed in\cite{steyn_roux_sasj_72}, and was applied for ZF SNR analysis in\cite[Refs.~24-27, 30, 31]{siriteanu_tvt_11}, as shown next.


\subsection{Ensuing Approximate Gamma Distributions for ZF SNRs Used in\cite[Refs.~24-27, 30, 31]{siriteanu_tvt_11} for $ r = \text{rank} (\uHd) = 1 $}
\label{section_approx_ZF_distr}

Based on~(\ref{equation_gamma_distribution_Gamma_exact_Rayleigh}), we can write for all the virtual ZF SNRs
\begin{eqnarray}
\label{equation_gammahat_Gamma}
&& \widehat{\gamma}_i = \frac{\Gamma_{\text{s}}}{\left[ \uWhatinv \right]_{i,i}} \sim {\cal{G}}  ( N, \widehat{\Gamma}_{K,i} ),  \widehat{\Gamma}_{K,i} = \frac{\Gamma_{\text{s}}}{\left[ \uRTKinvhat \right]_{i,i}}, \\
\label{equation_gammahat_Gamma_mgf}
&& M_{\widehat{\gamma}_i}(s) = \left( 1 - s \widehat{\Gamma}_{K,i}  \right)^{-N}, \, i = 1 : \NT. 
\end{eqnarray}
Then, the approximation in distribution $ \uW \,{\buildrel d \over \approx}\, \uWhat  $ from\cite{steyn_roux_sasj_72} led in\cite[Refs.~24-27, 30, 31]{siriteanu_tvt_11} to the approximation in distribution
\begin{eqnarray}
\label{equation_gammai_gammaihat_approx_distr}
\gamma_i \,{\buildrel d \over \approx}\, \widehat{\gamma}_i \sim {\cal{G}}  ( N, \widehat{\Gamma}_{K,i} ), \, i = 1 : \NT.
\end{eqnarray}
Finally, substituting the m.g.f.~from~(\ref{equation_gammahat_Gamma_mgf}) into~(\ref{equation_average_Pe}) has yielded the \textsl{approximate AEP expression}\cite[Eq.~(39)]{siriteanu_tvt_11}
\begin{eqnarray}
\label{equation_average_Pe_approximation}
\widehat{P}_{\text{e},i}^{(\ref{equation_average_Pe_approximation})} = \frac{1}{\pi} \int_{0}^{\frac{M-1}{M}\pi}
\left( 1 + \frac{ \sin^2{\frac{\pi}{M}}}{\sin^2 \theta} \widehat{\Gamma}_{K,i} \right)^{-N}\nid \theta, \, i = 1 : \NT.
\end{eqnarray}
The virtual SNR distribution~(\ref{equation_gammahat_Gamma}) and the ensuing $ \widehat{P}_{\text{e},i}^{(\ref{equation_average_Pe_approximation})} $ are referenced on Row 8 in Table~\ref{table_gamma1_statistics}, for virtual Rayleigh fading.


\subsection{Analogous Approximate CWD for $ \uGamma_1 $}
\label{section_analogous_SC}

Given $ v = 1 : \NT $, let us partition $ \uHhat $, $ \RThat $, $ \RThatinv $, $ \uWhat $, and $ \uWhatinv $ as done for $ \uH $, $ \uRTK $, $ \uRTKinv $, $ \uW $, and $ \uWinv $ in Section~\ref{section_system_channel_model}.
Also, analogously to the actual SC $ \uGamma_1 $ defined in~(\ref{equation_Schur_Complement_W}), let us define the virtual SC
\begin{eqnarray}
\label{equation_uGamma1hat_from_partition}
\widehat{\uGamma}_1 = (\uWhat^{11})^{-1} = \uWhat_{11} - \uWhat_{12} \uWhat_{22}^{-1} \uWhat_{21}.
\end{eqnarray}
Then, analogously to~(\ref{equation_gamma_from_W_G}), we can write
\begin{eqnarray}
\label{equation_gammai_hat_Gammahat}
\widehat{\gamma}_i = \frac{\Gamma_{\text{s}}}{\left[ \uWhatinv \right]_{i,i}} = \frac{\Gamma_{\text{s}}}{[\widehat{\uGamma}_1^{-1}]_{i,i}}, \, i = 1 : \NT.
\end{eqnarray}

Since $ \uHhat $ is zero-mean, $ \widehat{\uGamma}_1  $ has the m.g.f.
\begin{eqnarray}
\label{equation_uGamma1hat_mgf}
M_{\widehat{\uGamma}_1}(\uTheta) = {\big| \mbI_{v} - \uTheta \left( \RThatupaa \right)^{-1}  \big|^{-N_v}},
\end{eqnarray}
i.e., matrix $ \widehat{\uGamma}_1 $ has the following CWD:
\begin{eqnarray}
\label{equation_uGamma1hat_distribution}
\widehat{\uGamma}_1 \sim {\cal{CW}}_{v} \big(N_v, \big( \RThatupaa \big)^{-1} \big).
\end{eqnarray}
Based on the approximation in distribution $ \uW \,{\buildrel d \over \approx}\, \uWhat $ proposed in\cite{steyn_roux_sasj_72}, one may view $  \widehat{\uGamma}_1 $ as approximating $ \uGamma_1 $ in distribution.
This view is also supported by the fact that the generally-unknown distribution of $ \uGamma_1$ and the CWD of $ \widehat{\uGamma}_1 $ turn exactly the same under condition~(\ref{equation_special_condition_Hd1_Hd2_R21}), as shown next. 

\subsection{Condition for $ \uGamma_1  \,{\buildrel d \over =}\,  \widehat{\uGamma}_1 $, and for $ \gamma_i \,{\buildrel d \over =}\, \widehat{\gamma}_i $, $ i = 1 : v $}
\label{section_discussion_corroboration}

\begin{theorem}
\label{theorem_condition_makes_Wishart}
\begin{eqnarray}
\label{equation_uGamma1hat_distribution_central}
\uHda = \uHdb \uR_{2,1} \Leftrightarrow \left( \RThatupaa \right)^{-1} = (\RTupaa )^{-1}. 
\end{eqnarray}
\end{theorem}

\IEEEproof{
See Appendix~\ref{section_theorem_proof}.
}

\begin{corollary}
\label{corollary_condition_Gamma_Gammahat}
Theorems~\ref{theorem_Gamma1_central_condition} and~\ref{theorem_condition_makes_Wishart}, along with~(\ref{equation_uGamma1hat_distribution}), yield:
\begin{eqnarray}
\label{equation_uGamma1hat_distribution_central_1}
\uHda = \uHdb \uR_{2,1} \Leftrightarrow \widehat{\uGamma}_1 \,{\buildrel d \over =}\,  \uGamma_1 \sim  {\cal{CW}}_{v} \left( N_v, (\RTupaa )^{-1} \right). 
\end{eqnarray}
\end{corollary}

\begin{corollary}
\label{corollary_condition_exact_approx_gamma}
SNR--SC relationships from~(\ref{equation_gamma_from_W_G}) and~(\ref{equation_gammai_hat_Gammahat}) along with equivalence~(\ref{equation_uGamma1hat_distribution_central_1}) yield the implication
\begin{eqnarray}
\label{equation_gamma1_noncentral_v_1}
&& \uHda = \uHdb \uR_{2,1} \Rightarrow \forall i = 1 : v, \forall v = 1 : \NT  \nonumber \\
& & \gamma_i  = \frac{\Gamma_{\text{s}}}{[\uGamma_1^{-1}]_{i,i}} \,{\buildrel d \over =}\, \widehat{\gamma}_i  = \frac{\Gamma_{\text{s}}}{[\widehat{\uGamma}_1^{-1}]_{i,i}} \sim {\cal{G}}  \left( N, \Gamma_{K,i}  \right). \quad \quad 
\end{eqnarray}
\end{corollary}

Note that~(\ref{equation_gamma1_noncentral_v_1}) implies the AEP equality $ {P}_{\text{e},i}^{(\ref{equation_average_Pe_condition})}  = \widehat{P}_{\text{e},i}^{(\ref{equation_average_Pe_approximation})} $, $ i = 1 : v $, which is depicted in Rows 1--3 of Table~\ref{table_gamma1_statistics}.

\begin{corollary}
\label{corollary_condition_exact_approx_gamma_1}
For $ v = 1 $, i.e., scalar $ {\uGamma}_1 $ and $ \widehat{\uGamma}_1 $, equivalence~(\ref{equation_uGamma1hat_distribution_central_1}) yields  equivalence\footnote{Matrix $ \uR_{2,1} \doteq (\NT - v) \times v $ reduces to vector $ \ur_{2,1} \doteq (\NT - 1) \times 1 $.}
\begin{eqnarray}
\label{equation_gamma1_noncentral_central_same}
\uhda = \uHdb \ur_{2,1} \Leftrightarrow \gamma_1 \,{\buildrel d \over =}\, \widehat{\gamma}_1 \sim {\cal{G}}  \left( N, \Gamma_{K,1}  \right).
\end{eqnarray}
\end{corollary}

\subsection{Corroboration and Explanation of Previous Observations}
\label{section_corroborate}

The equivalence in~(\ref{equation_gamma1_noncentral_central_same}) helps explain our earlier observations that the accuracy of $ \gamma_1 \,{\buildrel d \over \approx }\, \widehat{\gamma}_1 $ and $ {P}_{\text{e},1}^{(\ref{equation_average_Pe_condition})} \approx  \widehat{P}_{\text{e},1}^{(\ref{equation_average_Pe_approximation})} $ is:
\begin{itemize}
\item dependent on the combination of $ \uHd $ and $ \uRTK $\cite[Sections~VI.B--E]{siriteanu_tvt_11}. 
\item poor for $ \uHd $ with rank $ r = \NT $\cite[Figs.~1, 2]{siriteanu_tvt_11} and for $ \uHd = \left( \begin{matrix} \uhda  & \mzero  \end{matrix} \right) $\cite[Figs.~1, 2]{siriteanu_twc_13}, whereby $ \uhda = \uHdb \ur_{2,1} $ does not hold.
\item good in\cite[Fig.~10]{siriteanu_twc_13} (and Fig.~\ref{figure_AEP_AER_vs_SNR_ZF_Rice_Ray_Fixed_AS_K_A1_NT3_NR4_Hd_6} herein) whereby $ \uhda \approx \uHdb \ur_{2,1} $.
\item inconsistent for $ r = 1 $\cite[Sections~VI.C]{siriteanu_tvt_11}, which is because $ r = 1 $ and $ \uhda = \uHdb \ur_{2,1}  $ are unrelated; thus, previous usage for $ r = 1 $  of $ \gamma_1 \,{\buildrel d \over \approx }\, \widehat{\gamma}_1 $ in\cite[Refs.~24-27,30,31]{siriteanu_tvt_11} appears unwarranted.
\end{itemize}

\section{M.G.F.~of matrix-SC $ \uGamma_1 $ under Rician$ (v) $/Rayleigh$(\NT - v)$ Fading, $  v = 1 : \NT - 1 $}
\label{section_uGamma1_distribution_Rician_Rayleigh_v_1}

\subsection{$ M_{\uGamma_1}(\uTheta) $ for $ \uHdb = \mzero $, in Terms of $ {_0\!F_0} \left( \uS,\uLambda \right)  $}
\label{section_mgf_Gamma1_general}

Our recent analysis of the scalar-SC (i.e., for $ v = 1 $) from\cite{siriteanu_twc_13} yielded for ZF under Rician$ (1) $/Rayleigh$(\NT - 1)$ fading the  SNR m.g.f.~for the (Rician) Stream 1 as\cite[Eq.~(31)]{siriteanu_twc_13}
\begin{eqnarray}
\label{equation_gamma1_mgf_final}
M_{\gamma_1}(s) = {\left( 1 - s \Gamma_{K,1} \right)^{-N}}{ {\,_1\!F_1}  \left(N; \NR; \sigma_1 \right)},
\end{eqnarray}
with $ \sigma_1 $ shown herein in Appendix~\ref{section_0_F_0_v_1}, Eq.~(\ref{equation_sigma1}), on page~\pageref{equation_sigma1}.
By substituting the confluent hypergeometric function of scalar argument from its infinite-series expansion around the origin\cite[Eq.~(13.2.2), p.~322]{NIST_book_10}
\begin{eqnarray}
\label{equation_1F1_definition}
{\,_1\!F_1} (N; \NR; \sigma_1) = \sum_{n = 0}^{\infty} { \frac{\left( N \right)_n}{\left( \NR \right)_n} \frac{\sigma_1^n}{n!} }
\end{eqnarray}
into~(\ref{equation_gamma1_mgf_final}), we showed in\cite[Eq.~(37)]{siriteanu_twc_13} that $ M_{\gamma_1}(s) $ is an infinite linear combination of m.g.f.s of Gamma distributions.


Hereafter, a matrix-SC analysis applicable $ \forall v = 1 : \NT - 1 $ characterizes the distribution of $ \uGamma_1 $ for Rician$ (v) $/Rayleigh$ (\NT - v) $ fading.
This analysis starts with the singular value decomposition
\begin{eqnarray}
\label{equation_H_U_Sigma_V}
\uHb = \uU \uSigma \uVH,
\end{eqnarray}
where $ \uU \doteq \NR \times \NR $, $ \uSigma \doteq \NR \times (\NT - v) $, and $ \uV \doteq (\NT - v) \times (\NT - v) $.
The unitary matrix $ \uU $, i.e., $ \uUH \uU = \uU \uUH = \mbI_{\NR} $, comprises the left singular vectors of $ \uHb $.
Using the definition of $ \uQ_2 $ from~(\ref{equation_Q_definition}) it can be shown that $ \uU $ is also the matrix with the eigenvectors of $ \uQ_2 $. Further, using~(\ref{equation_Projection_Q2}), we can write the eigendecomposition of $ \uQ_2 $ as:
\begin{eqnarray}
\label{equation_Q_U_Lambda_U}
\uQ_2 = \uUH \underbrace{ \text{diag} (\overbrace{1, \; 1, \; \cdots, \; 1}^{N_v}, \overbrace{0, \; 0, \; \cdots, \; 0}^{\NT-v} )}_{\uLambda \doteq \NR \times \NR} \uU.
\end{eqnarray}
Substituting~(\ref{equation_Q_U_Lambda_U}) into~(\ref{equation_uGamma1_mgf}) yields
\begin{eqnarray}
\label{equation_uGamma1_mgf_uU_1}
M_{\uGamma_1|\uU}(\uTheta) = {\big | \mbI_{v} - \uTheta \left( \RTupaa \right)^{-1} \big |^{-N_v}}  { \text{etr} \bigg( \overbrace{ \left[ \mbI_{v} - \uTheta \left( \RTupaa \right)^{-1}   \right]^{-1} \uTheta }^{= \uPsi } \uMH \uU \uLambda \uUH \uM  \bigg ) }. 
\end{eqnarray}

Now, averaging the $ \text{etr}(\cdot) $ term above over $ \uU $ appears to be tractable only for $ \uHdb = \mzero $, when matrix $ \uU $ has a known, Haar, distribution\cite{siriteanu_twc_13}.
This averaging has been pursued successfully for $ v = 1 $ in\cite{siriteanu_twc_13}.
Herein, we pursue, differently, the more general case whereby $ 1 \le v < \NT $.
Then,
\begin{eqnarray}
\label{equation_int_etr_SUTU}
\mathbb{E}_{\uU} \bigg\{  \text{etr} \left( \uPsi \uMH \uU \uLambda \uUH \uM  \right) \bigg\} 
& = & \int_{{\mathbb{U}}_{\NR}} {{\text{etr}} ( \underbrace{\uM \uPsi \uMH}_{=\uS} \uU \uLambda \uUH ) } [d \uU]  \nonumber \\ & = & \int_{{\mathbb{U}}_{\NR}} {\text{etr}} \left( \uS \uU \uLambda \uUH \right) [d \uU]  
= \int_{{\mathbb{U}}_{\NR}} \,{_0\!F_0} \left( \uS \uU \uLambda \uUH \right) [d \uU]. \nonumber
\end{eqnarray}
where $ {{\mathbb{U}}_{\NR}} $ is the unitary manifold comprising the $ \NR \times \NR $ unitary matrices with real diagonal elements, and $ [d \uU] $ is the normalized Haar invariant probability measure on $ {{\mathbb{U}}_{\NR}} $\cite[Appendix~1]{mckay_tcomm_09}. 
Matrix $ \uS \doteq \NR \times \NR $, which is given by
\begin{eqnarray}
\label{equation_S_definition}
\uS = \uM \uPsi \uMH = \uM \left[ \mbI_{v} - \uTheta \left( \RTupaa \right)^{-1}   \right]^{-1} \uTheta \uMH,
\end{eqnarray}
has rank $ v $ and distinct nonzero eigenvalues, in general.

Now, because\cite[Eq.~(92)]{james_ams_64}\cite[Eq.~(4.2)]{gross_jap_89}
\begin{eqnarray}
\label{equation_int_SUTU_0F0}
\int_{{\mathbb{U}}_{\NR}} \,{_0\!F_0} \left( \uS \uU \uLambda \uUH \right) [d \uU] = \,{_0\!F_0} \left( \uS,\uLambda \right), 
\end{eqnarray}
the m.g.f.~of the unconditioned $ \uGamma_1 $ can be written as
\begin{eqnarray}
\label{equation_uGamma1_mgf_uU_1_avg_1}
M_{\uGamma_1}(\uTheta)  ={\big | \mbI_{v} - \uTheta \left( \RTupaa \right)^{-1} \big |^{-N_v}} { \,{_0\!F_0} \left( \uS,\uLambda \right) }.
\end{eqnarray}

\subsection{Determinantal Expressions for $ {_0\!F_0} \left( \uS,\uLambda \right) $}
\label{section_determinantal_expressions_lists_0F0}

Appendix~\ref{section_uGamma1_distribution_Rician_Rayleigh_v} expresses $ \,{_0\!F_0} \left( \uS,\uLambda \right) $ as determinant of an $ \NR \times \NR $ matrix with elementary-function entries, as follows:



\begin{itemize}
\item in Appendix~\ref{section_distinct_eigenvalues}, from previous work{\cite{gross_jap_89}}{\cite{mckay_tcomm_09}}{\cite{chiani_tit_10}}, for when both $\uS$, $ \uLambda $ have nonequal eigenvalues, in  Eq.~(\ref{equation_int_SUTU}.
\item in Appendix~\ref{section_nondistinct_nondistinct}, for when both $ \uS $, $ \uLambda $ may have equal eigenvalues, in the new expression~(\ref{equation_Takemura_general_expression}).
\item in Appendix~\ref{section_S_nonfull_rank_L_idempotent}, for when $ \uS $ is rank-$ v $ with nonequal nonzero eigenvalues and $ \uLambda $ is idempotent and rank-$ N_v $ --- as for $ \uS $ from~(\ref{equation_S_definition}) and $ \uLambda $ from~(\ref{equation_Q_U_Lambda_U}), under Rician$ (v) $/Rayleigh$(\NT - v)$ fading --- in the new expression~(\ref{equation_denominator_rank_v_idempotent}).
Unfortunately, then,~(\ref{equation_uGamma1_mgf_uU_1_avg_1}) cannot yield SNR m.g.f.s based on the SNR--SC relationship from~(\ref{equation_gamma_from_W_G}) because the m.g.f.~of $ \uGamma_1^{-1} $ could not be deduced from~(\ref{equation_uGamma1_mgf_uU_1_avg_1}).
\item in Appendix~\ref{section_0_F_0_v_1}, for when $ \uS $ is rank-$ 1 $ and $ \uLambda $ is idempotent and rank-$ N $ --- as under Rician$ (1) $/Rayleigh$(\NT - 1)$ fading case --- in the new expression~(\ref{equation_int_etr_SUTU_S_r1_L_idem}). 
This case is considered in more detail below.

\end{itemize}

\subsection{New Determinantal Expressions for ZF SNR M.G.F.~and AEP~for Stream 1 Under Rician$ (1) $/ Rayleigh$(\NT - 1)$ Fading}
\label{section_determinantal_expressions_lists_0F0_1}

For Rician$ (1) $/Rayleigh$(\NT - 1)$ fading, partitioning with $ v = 1 $ has yielded in~(\ref{equation_gamma1_Gamma1}) the SNR-SC relationship $ \gamma_1 = \Gamma_{\text{s}} \uGamma_ 1 $, which, along with~(\ref{equation_uGamma1_mgf_uU_1_avg_1}), yields
\begin{eqnarray}
\label{equation_uGamma1_mgf_uU_1_avg_Nv_1}
M_{\gamma_1}(s) = M_{\uGamma_1}(s \Gamma_{\text{s}} )  = {( 1 - s \Gamma_{K,1} )^{-N}}{{{_0\!F_0}} \left( \uS,\uLambda \right)}.
\end{eqnarray}
Appendix~\ref{section_0_F_0_v_1} expressed $ {_0\!F_0} \left( \uS,\uLambda \right) $ for this case  in~(\ref{equation_int_etr_SUTU_S_r1_L_idem}) as 
\begin{eqnarray}
\label{equation_int_etr_SUTU_S_r1_L_idem_repeat}
{_0\!F_0} \left( \uS,\uLambda \right) = A \frac{ \Delta_2(N, \NR, \sigma_1)}{\sigma_1^{\NR - 1}},
\end{eqnarray}
where $ A $ is a scalar defined in~(\ref{equation_int_etr_SUTU_S_r1_L_idem}), and $ \Delta_2(N, \NR, \sigma_1) $ is the determinant of the matrix in~(\ref{equation_D3_S_1_L_idemp_num}).

Finally, substituting $ \sigma_1 $ from~(\ref{equation_sigma1}) into~(\ref{equation_int_etr_SUTU_S_r1_L_idem_repeat}), and the result
into~(\ref{equation_uGamma1_mgf_uU_1_avg_Nv_1}), yields for the Stream-1 SNR m.g.f.~the following new determinantal expression
\begin{eqnarray}
\label{equation_gamma1_mgf_final_123}
M_{\gamma_1}(s) = A \frac{\left( 1 - s \Gamma_{K,1} \right)^{\NT - 2}}{(s \Gamma_{K,1} \alpha )^{\NR - 1}} \Delta_2 \left( N, \NR, \frac{s \Gamma_{K,1}\alpha}{1-s \Gamma_{K,1} } \right),
\end{eqnarray}
with $\alpha$ defined in~(\ref{equation_S_rank_1}).
Then, substituting~(\ref{equation_gamma1_mgf_final_123}) into~(\ref{equation_average_Pe}) yields the corresponding new ZF {AEP expression} for Stream 1 under  Rician$ (1) $/Rayleigh$(\NT - 1)$ fading:
\begin{eqnarray}
\label{equation_average_Pe_Rice_Ray_determinant}
P_{\text{e},1}^{(\ref{equation_average_Pe_Rice_Ray_determinant})} = \frac{1}{\pi} \int_{0}^{\frac{M-1}{M}\pi}
A \frac{\left( 1 + {\frac{ \sin^2{\frac{\pi}{M}}}{\sin^2 \theta}} \Gamma_{K,1} \right)^{\NT - 2}}{ \left( {-\frac{ \sin^2{\frac{\pi}{M}}}{\sin^2 \theta}} \Gamma_{K,1} \alpha \right)^{\NR - 1}} \Delta_2 \left( N, \NR, \frac{-\Gamma_{K,1}\alpha\sin^2{\frac{\pi}{M}}}{\sin^2 \theta + \Gamma_{K,1} \sin^2{\frac{\pi}{M}} } \right)\nid \theta.
\end{eqnarray}
The SNR m.g.f.~expression~(\ref{equation_gamma1_mgf_final_123}) and the AEP expression~(\ref{equation_average_Pe_Rice_Ray_determinant}) are referenced in Table~\ref{table_gamma1_statistics}, Row 4, for Stream 1.

\begin{remark}
\label{remark_Ray_Rice_uncorrelated}
Under Rician$ (1) $/Rayleigh$(\NT - 1)$ fading, if the Rayleigh fading is uncorrelated with the Rician fading, the SNRs for the Rayleigh-fading streams, i.e., Streams $ i = 2 : \NT $, can be characterized by viewing this case as Rayleigh$(\NT - 1)$/Rician$ (1) $ fading that satisfies condition\footnote{The $ 1 \times (\NT - 1) $ vector $ \ur_{1,2}^{\cal{T}} $ is analogous of $ \uR_{2,1}  $ from~(\ref{equation_R21}).} $ \uHdb = \uhda \ur_{1,2}^{\cal{T}} = \mzero $. Based on Lemma~\ref{lemma_Gamma1_central_gamma_i_Gamma}, the SNRs for the Rayleigh-fading streams are then Gamma-distributed as in~(\ref{equation_gamma_distribution_Gamma_exact}), and their AEPs are described by expression~(\ref{equation_average_Pe_condition}) --- see Row 2 in Table~\ref{table_gamma1_statistics}.
Consequently, the AEP can then be computed for all streams: for the Rician-fading Stream 1 with~(\ref{equation_average_Pe_Rice_Ray_determinant}), and for the Rayleigh-fading streams with~(\ref{equation_average_Pe_condition}).
\end{remark}

\subsection{New Relationship between  $ {_0\!F_0} \left( \uS,\uLambda \right) $ and $ {_1\!F_1} \left(N; \NR; \sigma_1 \right) $, and Ensuing Determinantal Expression for $  {_1\!F_1} \left(N; \NR; \sigma_1 \right) $}
\label{section_determinantal_expressions_lists_0F0_relations}
\begin{corollary}
If $ \uS $ and $ \uLambda $ are $ \NR \times \NR $ matrices, $ \uS $ of rank $ 1 $ with nonzero eigenvalue $ \sigma_1 $ and $ \uLambda $ idempotent of rank $ N $ then~(\ref{equation_gamma1_mgf_final}) and~(\ref{equation_uGamma1_mgf_uU_1_avg_Nv_1}) reveal the previously unknown relationship
\begin{eqnarray}
\label{equation_int_etr_SUTU_1F1}
{_0\!F_0} \left( \uS,\uLambda \right) = {_1\!F_1} \left(N; \NR; \sigma_1 \right).
\end{eqnarray}
\end{corollary}

\begin{corollary}
Eqs.~(\ref{equation_int_etr_SUTU_1F1}) and~(\ref{equation_int_etr_SUTU_S_r1_L_idem_repeat}) yield for $ {\,_1\!F_1} \left(N; \NR; \sigma_1 \right) $ the  new --- determinantal --- expression
\begin{eqnarray}
\label{equation_1F1_det_expression}
{\,_1\!F_1}  \left(N; \NR; \sigma_1 \right) = A { \Delta_2(N, \NR, \sigma_1)}/{\sigma_1^{\NR - 1}}.
\end{eqnarray}
\end{corollary}

\section{Numerical Results}
\label{section_Numerical_Results}

\subsection{Description of Settings}
\label{section_Numerical_Results_Settings}

For $ v = 1 $, i.e., the partitioning from~(\ref{equation_partitioned_H_general_v_1}), Stream-1 AEP results obtained in \texttt{MATLAB} are presented for $ \NR = 4 $, $ \NT = 3 $, QPSK modulation, and relevant ranges of the average SNR per transmitted bit $ \Gamma_{\text{b}} = \frac{ \Gamma_{\text{s}} }{\log_2 M} $. 
Matrix $ \uRT $ has been computed as in\cite{siriteanu_tvt_11}, for a uniform linear antenna array with interelement distance normalized to carrier half-wavelength $ d_{\text{n}} = 1 $, Laplacian power azimuth spectrum centered at $ \theta_{\text{c}} = 5^\circ $, and $ K $ and AS set to their lognormal-distribution averages for two WINNER II scenarios\cite[Table~I]{siriteanu_tvt_11}: 
\begin{itemize}
\item B1 (typical urban microcell): $ K = 9 $~dB, $ \text{AS} = 3^{\circ} $, i.e., high transmit-correlation, and, thus, $ \ur_{2,1} \neq \mzero $.
\item A1 (indoor office/residential): $ K = 7 $~dB, $ \text{AS} = 51^{\circ} $, i.e., low correlation, and, thus, $ \ur_{2,1} \approx \mzero $.
\end{itemize}
For consistency with our previous work in\cite{siriteanu_tvt_11}\cite{siriteanu_twc_13}, results are shown herein for $ v = 1 $, $ [\RT]_{i,i} $, $ i = 1 : \NT $, i.e., for $ \mathbb{E} \{ | [\uH_{\text{r,norm}}]_{i,j}  |^2 = 1 \} $, $ \forall i, j $, and $ \uH_{\text{d,norm}} $ with arbitrary complex-valued elements\footnote{$ \RT $ and $ \uH_{\text{d,norm}} $ were adjusted to satisfy~(\ref{equation_special_condition_Hd1_Hd2_R21}), when necessary.}. 
Nevertheless, other (unshown) results have validated our analysis against simulations also for $ v > 1 $, and $ \RT $ and $ \uH_{\text{d,norm}} $ generated as for a MIMO system with distributed transmitters, based on\cite{chen_tsp_10}.

In our figures, the legends identify results from exact and approximate AEP expressions (with \texttt{exact}, \texttt{approx}) and from simulation of $ 10^6$ channel and noise samples (with \texttt{sim}).
All figures depict Rayleigh--only fading, with red lines and markers, and with legend \texttt{Ray}--\texttt{Ray}.
Additionally, each figure depicts, with black lines and markers, one of the following Rician-fading cases: full-Rician (\texttt{Rice}--\texttt{Rice}), Rayleigh--Rician (\texttt{Ray}--\texttt{Rice}), or Rician--Rayleigh (\texttt{Rice}--\texttt{Ray}), for $ \uhda = \uHdb \ur_{2,1} $, $ \uhda \approx \uHdb \ur_{2,1} $, or $ \uhda \neq \uHdb \ur_{2,1} $.
Each case is also identified in figures and discussion by the corresponding row number in Table~\ref{table_gamma1_statistics}. 


\subsection{Full-Rician Fading, High Correlation, $ \uhda = \uHdb \ur_{2,1} $}
\label{section_Numerical_Results_Rice_rice_cond_ok}

\begin{figure}[t]
\begin{center}
\includegraphics[width=4.0in]
{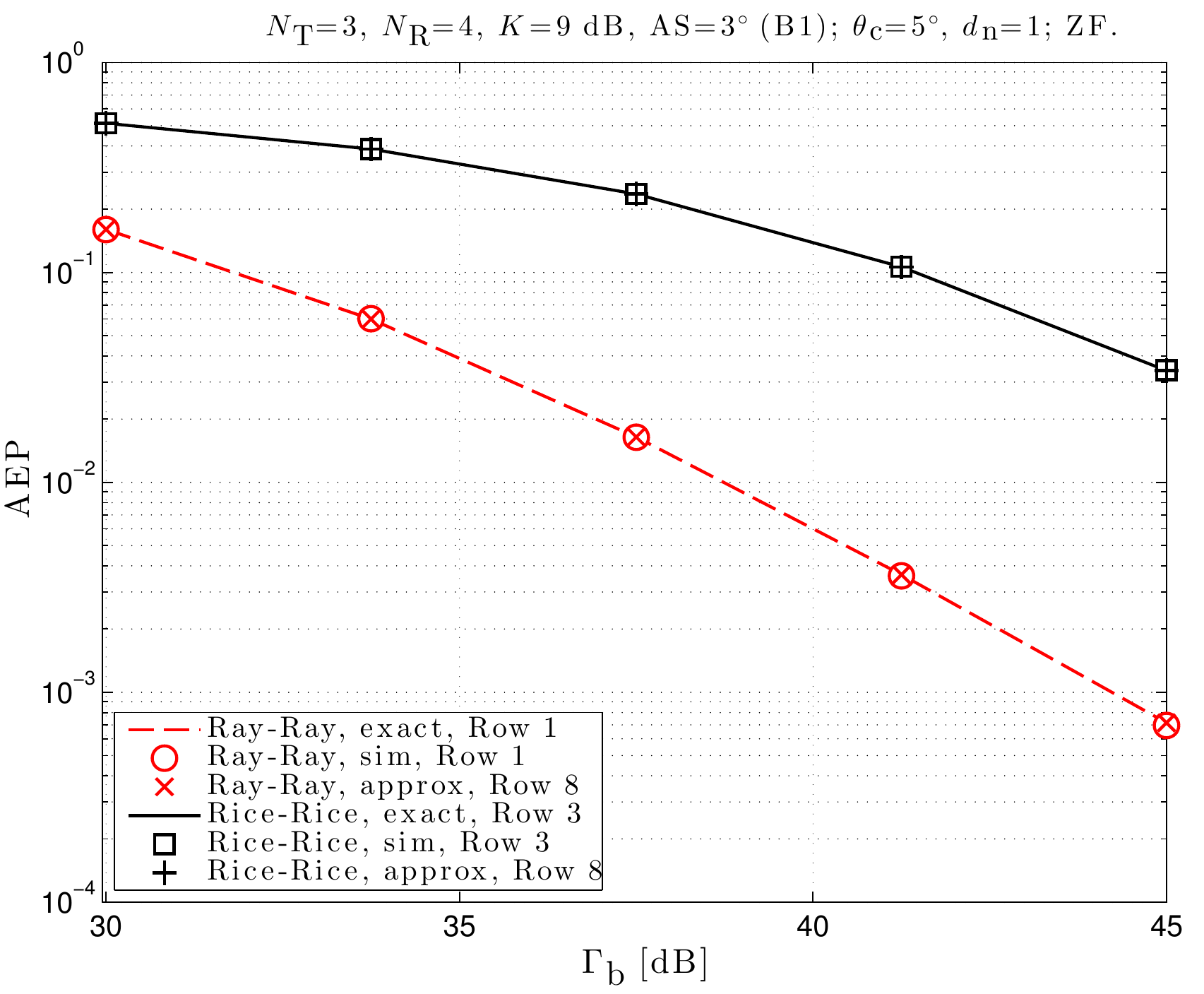}
\caption{Stream-1 AEP from exact expression~(\ref{equation_average_Pe_condition}), approximate expression~(\ref{equation_average_Pe_approximation}), and from simulation, for Rayleigh-only fading and for full-Rician fading under condition $ \uhda = \uHdb \ur_{2,1} $, for QPSK modulation, $ \NR = 4 $, $ \NT = 3 $, $ K = 9 $~dB, $ \text{AS} = 3^{\circ} $ (i.e., WINNER II scenario B1 averages).}
\label{figure_AEP_AER_vs_SNR_ZF_Rice_Ray_Fixed_AS_K_B1_NT3_NR4_Hd_3}
\end{center}
\end{figure}

Fig.~\ref{figure_AEP_AER_vs_SNR_ZF_Rice_Ray_Fixed_AS_K_B1_NT3_NR4_Hd_3} depicts full-Rician fading, i.e., $ \uhda \neq \mzero $, and $ \uHdb \neq \mzero $, under condition $ \uhda = \uHdb \ur_{2,1} $, which is characterized in Row 3, for scenario B1.
Note first that analysis and simulation results agree.
Then, as predicted by Corollary~\ref{corollary_condition_exact_approx_gamma_1}, the AEP from the exact and approximate expressions agree, because $ \uhda = \uHdb \ur_{2,1} $.
Finally, as predicted by Corollary~\ref{corollary_Ray_Rice}, Rician fading yields poorer performance than Rayleigh-only fading.

\subsection{Rayleigh--Rician, High Correlation, i.e., $ \uhda \neq \uHdb \ur_{2,1} $}
\label{section_Numerical_Results_Ray_Rice_cond_not_ok}

\begin{figure}[t]
\begin{center}
\includegraphics[width=4.0in]
{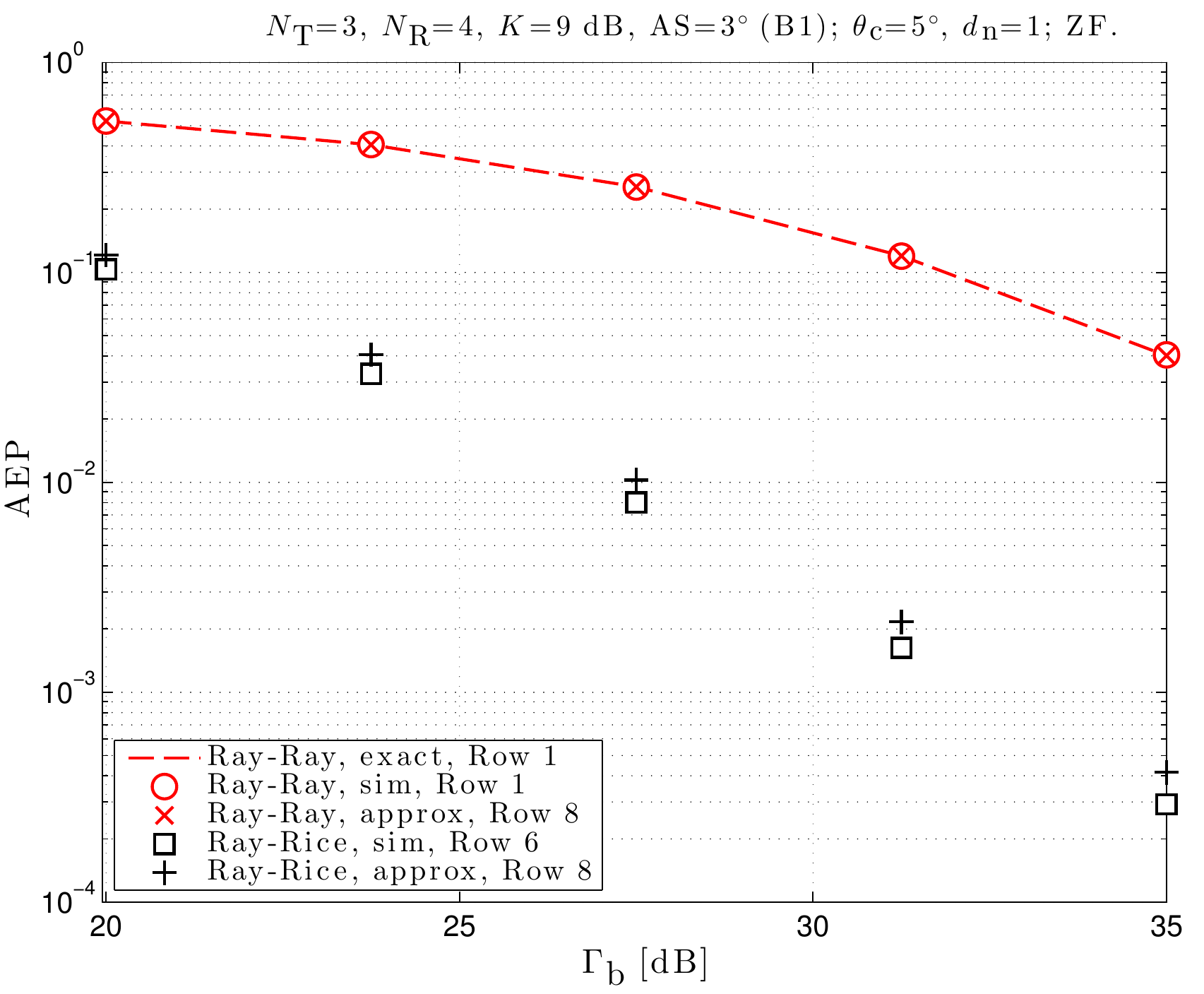}
\caption{Stream-1 AEP from exact expression~(\ref{equation_average_Pe_condition}), approximate expression~(\ref{equation_average_Pe_approximation}), and from simulation, for Rayleigh-only fading and for Rayleigh$ (1) $/Rician$(\NT - 1)$ fading under conditions $ \uhda = \mzero $ and $ \uHdb \neq \mzero $, for QPSK modulation, $ \NR = 4 $, $ \NT = 3 $, $ K = 9 $~dB, $ \text{AS} = 3^{\circ} $ (i.e., WINNER II scenario B1 averages). Since $ \ur_{2,1} \neq \mzero $,  we have $ \uhda \neq \uHdb \ur_{2,1} $.}
\label{figure_AEP_AER_vs_SNR_ZF_Rice_Ray_Fixed_AS_K_B1_NT3_NR4_Hd_6}
\end{center}
\end{figure}

Fig.~\ref{figure_AEP_AER_vs_SNR_ZF_Rice_Ray_Fixed_AS_K_B1_NT3_NR4_Hd_6} depicts Rayleigh$ (1) $/Rician$(\NT - 1)$ fading with $ \uhda = \mzero $, $ \uHdb \neq \mzero $, for scenario B1, i.e., $ \ur_{2,1} \neq \mzero $, so that $ \uhda \neq \uHdb \ur_{2,1} $, which is characterized in Row 6.
Since no exact AEP expression is then known, Fig.~\ref{figure_AEP_AER_vs_SNR_ZF_Rice_Ray_Fixed_AS_K_B1_NT3_NR4_Hd_6} shows results only from simulation and approximation (see the \texttt{Ray}--\texttt{Rice} plots with black $ \square $ and $ + $ markers), which do not agree because $ \uhda \neq \uHdb \ur_{2,1} $.
Further, the plots with black $ \square $ and red  $ \ocircle $ markers reveal a surprising phenomenon for this fading case, i.e., when the intended stream undergoes Rayleigh fading that is highly-correlated with the interfering fading: Rician-fading {interference} yields much better performance than Rayleigh-fading interference. 


\subsection{Rayleigh--Rician, Low Correlation, i.e., $ \uhda \approx \uHdb \ur_{2,1} $}
\label{section_Numerical_Results_Ray_Rice_cond_approx}

\begin{figure}[t]
\begin{center}
\includegraphics[width=4.0in]
{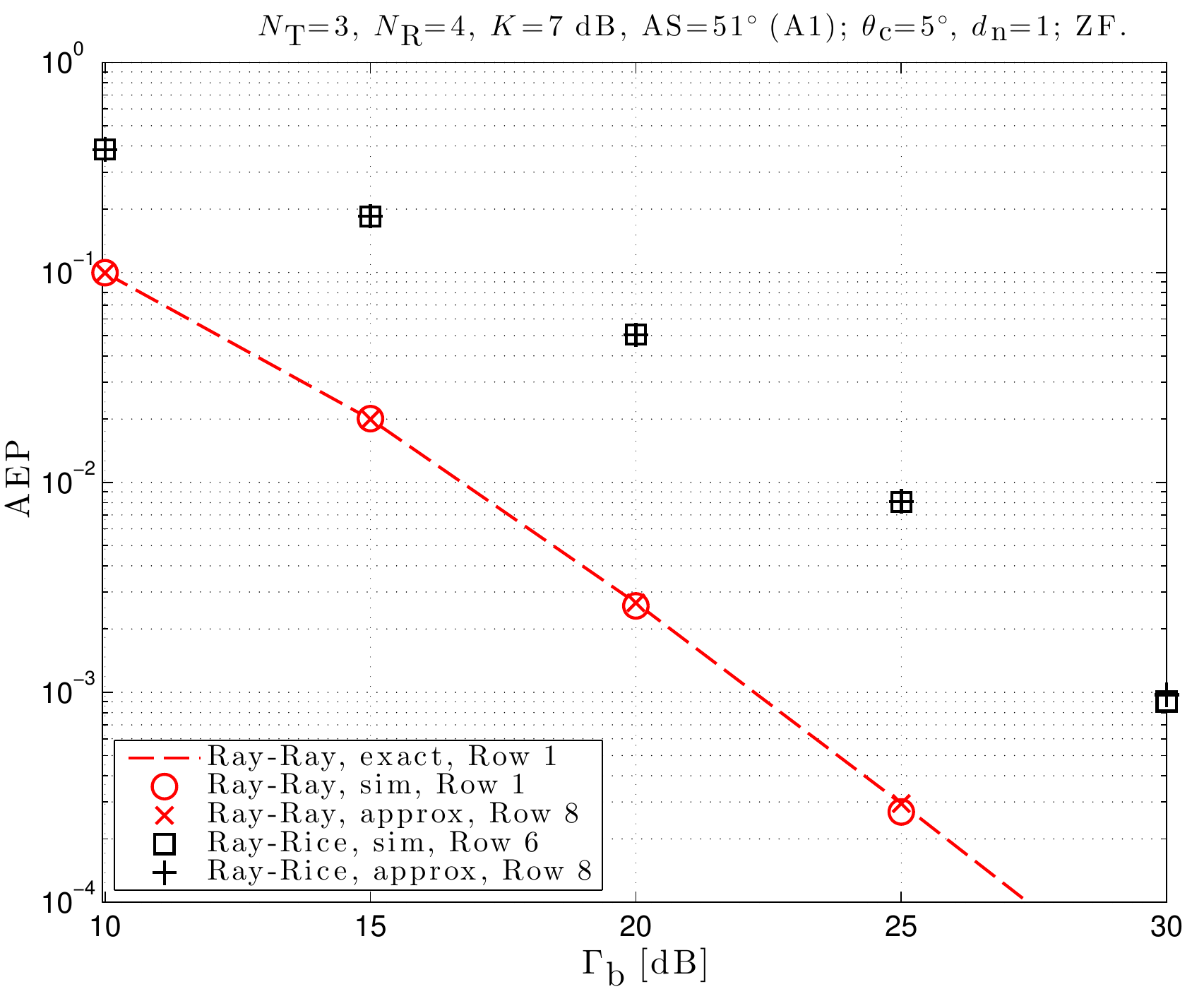}
\caption{Stream-1 AEP from exact expression~(\ref{equation_average_Pe_condition}), approximate expression~(\ref{equation_average_Pe_approximation}), and from simulation, for Rayleigh-only fading and for Rayleigh$ (1) $/Rician$(\NT - 1)$ fading under conditions $ \uhda = \mzero $ and $ \uHdb \neq \mzero $, for QPSK modulation, $ \NR = 4 $, $ \NT = 3 $, $ K = 7 $~dB, $ \text{AS} = 51^{\circ} $ (i.e., WINNER II scenario A1 averages). Since $ \ur_{2,1} \approx \mzero $,  we have $ \uhda \approx \uHdb \ur_{2,1} $.}
\label{figure_AEP_AER_vs_SNR_ZF_Rice_Ray_Fixed_AS_K_A1_NT3_NR4_Hd_6}
\end{center}
\end{figure}

Fig.~\ref{figure_AEP_AER_vs_SNR_ZF_Rice_Ray_Fixed_AS_K_A1_NT3_NR4_Hd_6} depicts the same fading cases as Fig.~\ref{figure_AEP_AER_vs_SNR_ZF_Rice_Ray_Fixed_AS_K_B1_NT3_NR4_Hd_6}, but for scenario A1, i.e., for low correlation \footnote{We obtained similar results in\cite[Fig.~10]{siriteanu_twc_13}.}.
This yields $ \ur_{2,1} \approx \mzero $ and, because $ \uhda = \mzero $, we have\footnote{
This case is characterized, approximately, by Row 2.} $ \uhda \approx \uHdb \ur_{2,1} $, which explains the agreement between the AEP from simulation and the approximate expression for the \texttt{Ray}--\texttt{Rice} plots (black markers).
Unshown results have confirmed that, for Rayleigh$ (1) $/Rician$(\NT - 1)$ fading, the approximate and exact distributions of ZF SNR for Stream 1 become more similar with less correlation between the Rayleigh and Rician fading.

Further, the plots with black $ \square $ and red $ \ocircle $ markers reveal the following for this fading case, i.e., when the intended stream undergoes Rayleigh fading that is nearly uncorrelated with the interfering fading: Rician-fading {interference} yields poorer performance than Rayleigh-fading interference, as predicted by Corollary~\ref{corollary_Ray_Rice} and Remark~\ref{remark_M_zero}, as $ \uhda \approx \uHdb \ur_{2,1} $. 


\subsection{Rician--Rayleigh, Low Correlation, $ \uhda \neq \uHdb \ur_{2,1} $}
\label{section_Numerical_Results_Rice_Ray_cond_not_ok}

\begin{figure}[t]
\begin{center}
\includegraphics[width=4.0in]
{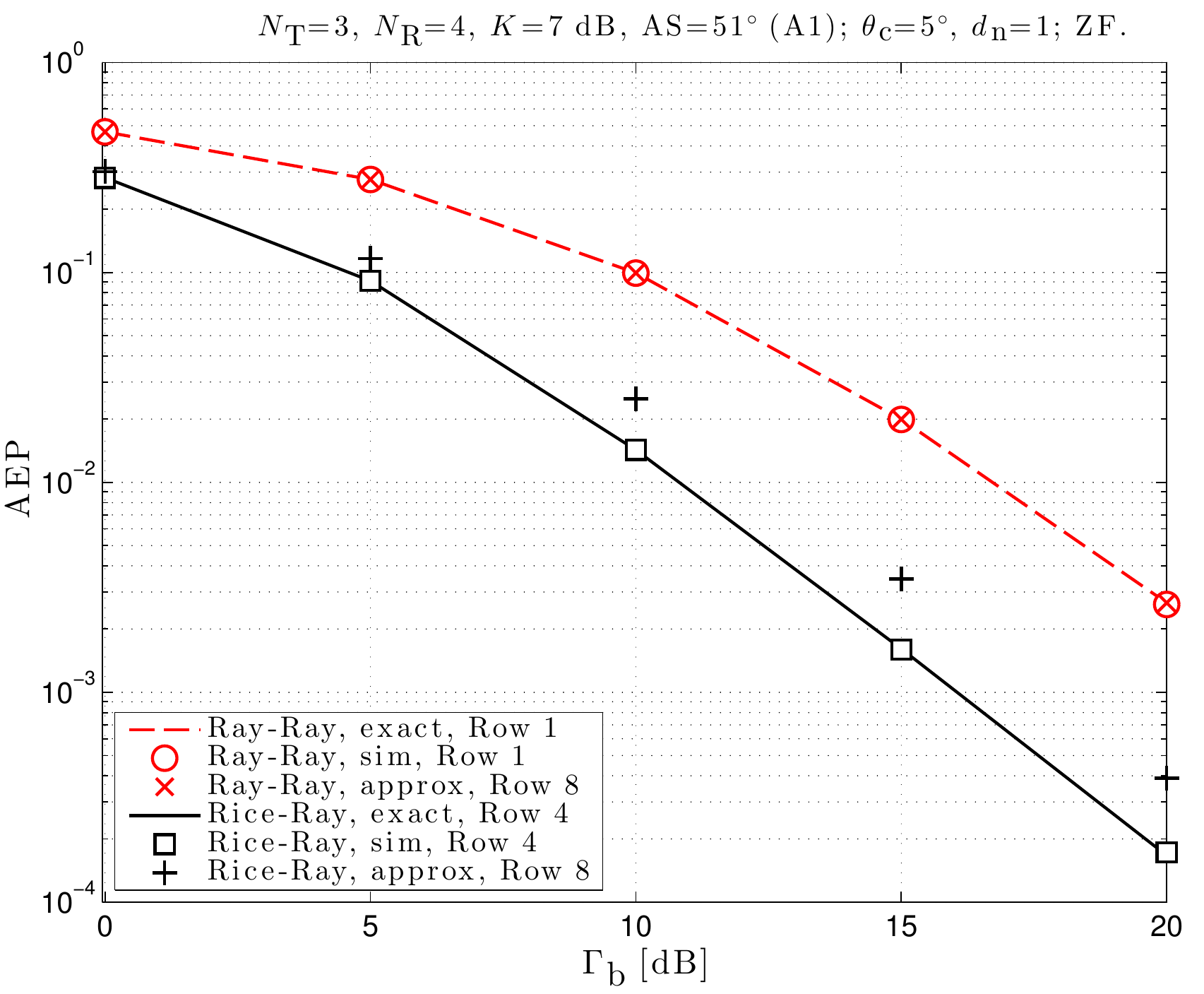}
\caption{Stream-1 AEP from exact expressions~(\ref{equation_average_Pe_condition}) and~(\ref{equation_average_Pe_Rice_Ray_determinant}), approximate expression~(\ref{equation_average_Pe_approximation}), and from simulation, for Rayleigh-only fading and for Rician$ (1) $/Rayleigh$(\NT - 1)$ fading under conditions $ \uhda \neq \mzero $ and $ \uHdb = \mzero $, i.e., $ \uhda \neq \uHdb \ur_{2,1} $, for QPSK modulation, $ \NR = 4 $, $ \NT = 3 $, $ K = 7 $~dB, $ \text{AS} = 51^{\circ} $ (i.e., WINNER II scenario A1 averages).}
\label{figure_AEP_AER_vs_SNR_ZF_Rice_Ray_Fixed_AS_K_A1_NT3_NR4_Hd_4}
\end{center}
\end{figure}

Fig.~\ref{figure_AEP_AER_vs_SNR_ZF_Rice_Ray_Fixed_AS_K_A1_NT3_NR4_Hd_4} depicts Rician$ (1) $/Rayleigh$(\NT - 1)$ fading, i.e., $ \uhda \neq \mzero $ and $ \uHdb = \mzero $, for scenario A1. 
This case implies $ \uhda \neq \uHdb \ur_{2,1} $, and is characterized in Row 4.
The new exact determinantal AEP expression~(\ref{equation_average_Pe_Rice_Ray_determinant}) agrees with the simulation results, but not with the approximate AEP expression~(\ref{equation_average_Pe_approximation}), which is explained by Corollary~\ref{corollary_condition_exact_approx_gamma_1}, as $ \uhda \neq \uHdb \ur_{2,1} $.



\subsection{Condition $ \uhda = \uHdb \ur_{2,1} $ Impact on Relative Performance}

The relative positions of plots with black vs.~red lines in the figures reveal that if condition $ \uhda = \uHdb \ur_{2,1} $ holds then Rayleigh-only fading outperforms Rician fading, e.g., in Figs.~\ref{figure_AEP_AER_vs_SNR_ZF_Rice_Ray_Fixed_AS_K_B1_NT3_NR4_Hd_3} and~\ref{figure_AEP_AER_vs_SNR_ZF_Rice_Ray_Fixed_AS_K_A1_NT3_NR4_Hd_6} --- which is supported by Corollary~\ref{corollary_Ray_Rice} and Remark~\ref{remark_M_zero} --- and also also \textit{vice versa}, e.g., in Figs.~\ref{figure_AEP_AER_vs_SNR_ZF_Rice_Ray_Fixed_AS_K_B1_NT3_NR4_Hd_6} and~\ref{figure_AEP_AER_vs_SNR_ZF_Rice_Ray_Fixed_AS_K_A1_NT3_NR4_Hd_4}.

\section{Summary and Conclusions}
\label{section_conclusions}
By characterizing the distribution of the matrix-SC in the NCWD Gramian matrix induced by a nonzero-mean Gaussian matrix, we analyzed MIMO ZF under transmit-correlated Rician fading.
Although expressing the m.g.f.~of the unconditioned matrix-SC (and the ZF SNRs) remains intractable for general Rician fading, we have succeeded for two cases.

The first tractable case arose by imposing the mean--correlation condition that yields a CWD for the matrix-SC.
We have shown that this condition also renders exact a previously-proposed approximation with the Gamma distribution of the unknown distribution of the ZF SNRs under Rician fading.
This finding has corroborated previous observations made in our work, and explained accuracy inconsistencies observed for the fading case usually assumed by others.

The second tractable case is that of Rician--Rayleigh fading.
Then, for the matrix-SC m.g.f., we have derived new expressions in terms of the determinant of a matrix with elementary-function entries.
Thus, we have also obtained new, determinantal expressions for the ZF SNR m.g.f.~and AEP.
Finally, we have revealed new determinantal expressions for, and a new relationship between, hypergeometric functions of matrix and scalar arguments.

Numerical results have confirmed analysis predictions, i.e., that: 1) the previously-proposed approximation becomes exact under the newly-discovered condition; 2) Rician-fading streams may still experience Rayleigh-like SNR distributions, under the newly-discovered condition; 3) the condition also determines the relative performance with Rician vs.~Rayleigh-only fading. 
Finally, numerical results have  also revealed a surprising phenomenon  when the intended stream undergoes Rayleigh fading and the intended and interfering fading are highly correlated: Rician-fading interference can then greatly benefit performance vs.~Rayleigh-fading interference.


\begin{appendices}

\section{Further Results on Condition $ \uHda = \uHdb \uR_{2,1}$}
\label{section_further_corollaries}

Recall that, for $ \uH \sim {\cal{CN}} \left(\uHd, \mbI_{\NR} \otimes \uRTKH \right) $,  $ \uRTK  $ is the covariance matrix of the columns of $ \uHH $. 
Using the UL decomposition of $ \uRTK = \uA \uAH $, and defining $ \uHw \sim {\cal{CN}} \left(\mzero, \mbI_{\NR} \otimes \mbI_{\NT} \right) $, we can write
\begin{eqnarray}
\uH = \uHd + \uHw \uAH,
\end{eqnarray}
so that
\begin{eqnarray}
\label{equation_HA}
\uH \uAinvH  = \uHd \uAinvH + \uHw.
\end{eqnarray}
Based on the partitionings of $ \uHd $ and $ \uAinvH $, we can write
\begin{eqnarray}
  \label{equation_Hd_A_our}
\uHd \uAinvH = \left( \begin{matrix}
   \uHda  \uA^{11,{\cal{H}}}  + \uHdb \uA^{12,{\cal{H}}} & \uHdb \uA^{22,{\cal{H}}} \\
  \end{matrix}
  \right), \nonumber
\end{eqnarray}
which, based on~(\ref{equation_A21A11A22A21_our})--(\ref{equation_RTK11inv_our}) and~(\ref{equation_R21}), becomes
\begin{eqnarray}
\label{equation_partition_Hd_AminusH}
\uHd \uAinvH = \left( \begin{matrix}
    \left[ \uHda - \uHdb \uR_{2,1} \right] \uA^{11,{\cal{H}}} & \uHdb  \uA^{22,{\cal{H}}} 
  \end{matrix}
  \right).
\end{eqnarray}
Finally,~(\ref{equation_special_condition_Hd1_Hd2_R21}),~(\ref{equation_HA}), and~(\ref{equation_partition_Hd_AminusH}) prove the following Lemma.

\begin{lemma}
\label{lemma_special_rel_makes_zero_our}
\begin{eqnarray}
\label{equation_condition_effect_on_Hd_our}
&& \uHda = \uHdb \uR_{2,1} \Leftrightarrow \uH \uAinvH = \left(  
  \begin{matrix}
        \mzero & \uHdb  \uA^{22,{\cal{H}}}
  \end{matrix}
  \right) +
  \left(  
  \begin{matrix}
        \uHwa & \uHwb
  \end{matrix}
  \right)
\end{eqnarray}
i.e., the mean--correlation condition is equivalent with the fact that canceling the transmit-correlation in the channel matrix yields a matrix whose first $ v $ columns are zero-mean.
\end{lemma}

The following corollary summarizes from Theorem~\ref{theorem_Gamma1_central_condition} and Lemma~\ref{lemma_special_rel_makes_zero_our} the necessary and sufficient conditions for $ \uGamma_1 $ to be CWD.

\begin{corollary}
\label{corollary_condition_new_cond_wishart}
\begin{eqnarray}
&& \uGamma_1 \sim  {\cal{CW}}_{v} \left(N_v, \left( \RTupaa \right)^{-1} \right) \Leftrightarrow \uHda = \uHdb \uR_{2,1}   \Leftrightarrow \uH \uAinvH = \left(  
  \begin{matrix}
        \mzero & \uHdb  \uA^{22,{\cal{H}}}
  \end{matrix}
  \right) +
  \left(  
  \begin{matrix}
        \uHwa & \uHwb
  \end{matrix}
  \right). \nonumber
\end{eqnarray}
\end{corollary}

\begin{corollary}[{Mean--Correlation `Parallelism'}]
\label{corollary_parallelism}
For nonsingular $ \uHdbH \uHdb $,
\begin{eqnarray}
\label{equation_ua_h1_Htilde_1}
\uHda = \uHdb \uR_{2,1} \Rightarrow \left( \uHdbH \uHdb \right)^{-1} \left( \uHdbH \uHda \right) =  \left( \mathbb{E} \{  \uHrbH \uHrb\}  \right)^{-1} \left( \mathbb{E} \{\uHrbH \uHra  \} \right). 
\end{eqnarray}
\end{corollary}

\IEEEproof{Follows by premultiplying $ \uHda = \uHdb \RTbbinva \RTba $ with $ \left( \uHdbH \uHdb \right)^{-1} \uHdbH $,
and expressing $ \RTbb $ and $ \RTba $ from~(\ref{equation_RTK_Hr}).}


\section{Proof of Theorem~\ref{theorem_condition_makes_Wishart}: {$ \uHda = \uHdb \uR_{2,1} \Leftrightarrow \left( \RThatupaa \right)^{-1} = \left( \RTupaa \right)^{-1} $}}
\label{section_theorem_proof}

Let us first find a simpler condition equivalent with $ \left( \RThatupaa \right)^{-1} = \left( \RTupaa \right)^{-1} $.
Equalizing the SC representation for $ (\RTupaa )^{-1} $ from~(\ref{equation_Schur_Complement_RT}) with that obtained analogously for $ ( \RThatupaa )^{-1}  $ based on~(\ref{equation_virtual_correlation_matrix}) yields
\begin{eqnarray}
&&\cancel{\RTaa} - \RTab \RTbbinva \RTba = \cancel{\RTaa} + \frac{1}{\NR} \uHdaH \uHda \nonumber \\ 
&& \quad - (\RTab + \frac{1}{\NR} \uHdaH \uHdb)\underbrace{(\RTbb + \frac{1}{\NR} \uHdbH \uHdb )^{-1}}_{=\uP}   (\RTba + \frac{1}{\NR} \uHdbH \uHda ), 
\end{eqnarray}
i.e.,
\begin{eqnarray}
&& \frac{1}{\NR} \uHdaH \uHda + \RTab \RTbbinva \RTba = (\RTab + \frac{1}{\NR} \uHdaH \uHdb) \uP (\RTba + \frac{1}{\NR} \uHdbH \uHda ), \nonumber
\end{eqnarray}
or
\begin{eqnarray}
\label{equation_Riccati_11} 
&& \uHdaH \overbrace{ \left( \mbI_{\NR} - \frac{1}{\NR} \uHdb \uP \uHdbH \right) }^{=\uQ} \uHda  + \NR \RTab \left( \RTbbinva - \uP \right) \RTba \nonumber \\ && \quad \; = \uHdaH \overbrace{ \uHdb \uP \RTba}^{=\uF}  + \overbrace{\RTab \uP\uHdbH }^{=\uFH} \uHda, \nonumber
\end{eqnarray}
or, finally,
\begin{eqnarray}
\label{equation_Riccati_112} 
&& \uHdaH \uQ \uHda + \NR \RTab \left( \RTbbinva - \uP \right) \RTba = \uHdaH \uF + \uFH\uHda.
\end{eqnarray}
The Woodbury matrix-inversion formula\cite[p.~165]{zhang_book_05} yields
\begin{eqnarray}
\label{equation_Riccati_1_Q}
\uQ & = & \mbI_{\NR} - \frac{1}{\NR} \uHdb \left( \RTbb + \frac{1}{\NR} \uHdbH \uHdb \right)^{-1} \uHdbH = \left( \mbI_{\NR} + \frac{1}{\NR} \uHdb \RTbbinva \uHdbH \right)^{-1} \\
\label{equation_Riccati_1_R_minus_P}
\uP & = & \left( \RTbb + \frac{1}{\NR} \uHdbH \uHdb \right)^{-1}\nonumber \\
& = & \RTbbinva - \RTbbinva \uHdbH  \frac{1}{\NR} \left( \mbI_{\NR} + \frac{1}{\NR} \uHdb \RTbbinva \uHdbH \right)^{-1} \uHdb \RTbbinva, \nonumber
\end{eqnarray}
i.e., 
\begin{eqnarray}
\label{equation_R_minus_P}
\RTbbinva - \uP = \frac{1}{\NR} \RTbbinva \uHdbH \uQ \uHdb \RTbbinva.
\end{eqnarray}
Substituting~(\ref{equation_R_minus_P}) into~(\ref{equation_Riccati_112}) yields
\begin{eqnarray}
\label{equation_Riccati_2}
&& \uHdaH \uQ  \uHda + \overbrace{\RTab \RTbbinva \uHdbH}^{=\uBH} \uQ \overbrace{\uHdb \RTbbinva \RTba}^{=\uB} = \uHdaH \uF + \uFH \uHda, \nonumber
\end{eqnarray}
or
\begin{eqnarray}
\label{equation_Riccati_5}
\uHdaH \uQ  \uHda - \uHdaH \uF - \uFH \uHda + \uBH \uQ \uB  = \mzero,
\end{eqnarray}
where
\begin{eqnarray}
\label{equation_Riccati_4}
\uF & = & \uHdb \uP \RTba = \uHdb \! \left( \! \RTbbinva - \frac{1}{\NR} \RTbbinva \uHdbH \uQ \uHdb \RTbbinva \! \right) \! \RTba \nonumber \\
& = & \uB - \underbrace{\frac{1}{\NR}\uHdb \RTbbinva \uHdbH}_{\,{\buildrel (\ref{equation_Riccati_1_Q}) \over =}\, \uQ^{-1} - \mbI_{\NR}} \uQ \uB = \uB - (\uQ^{-1} - \mbI_{\NR}) \uQ \uB = \uQ \uB. \nonumber
\end{eqnarray}
Thus,~(\ref{equation_Riccati_5}) becomes
\begin{eqnarray}
\label{equation_Riccati_6}
\uHdaH \uQ  \uHda - \uHdaH \uQ \uB - \uBH \uQ \uHda + \uBH \uQ \uB = \mzero,
\end{eqnarray}
which is the sought simpler expression equivalent with $ \left( \RThatupaa \right)^{-1} = \left( \RTupaa \right)^{-1} $.

Now, let us assume that $ \left( \RThatupaa \right)^{-1} = \left( \RTupaa \right)^{-1} $ holds, i.e., that~(\ref{equation_Riccati_6}) holds.
Then, with $ \uHdatilde = \uQ^{1/2} \uHda $ and $ \uBtilde = \uQ^{1/2} \uB $,~(\ref{equation_Riccati_6}) becomes
\begin{eqnarray}
\label{equation_Riccati_7}
\uHdatildeH \uHdatilde - \uHdatildeH \uBtilde - \uBtildeH \uHdatilde + \uBtildeH \uBtilde = \mzero,
\end{eqnarray}
which can be written further as
\begin{eqnarray}
\label{equation_Riccati_8}
\uHdatildeH \left( \uHdatilde - \uBtilde \right) - \uBtildeH \left( \uHdatilde - \uBtilde \right) = \mzero,
\end{eqnarray}
or
\begin{eqnarray}
\label{equation_Riccati_88}
\left( \uHdatilde - \uBtilde \right)^{\cal{H}} \left( \uHdatilde - \uBtilde \right) = \mzero,
\end{eqnarray}
which implies
\begin{eqnarray}
\label{equation_Riccati_9}
\uHdatilde =  \uBtilde \Leftrightarrow \uHda = \uB = \uHdb \RTbbinva \RTba = \uHdb \uR_{2,1}. \nonumber
\end{eqnarray}

Assuming, conversely, that $ \uHda  =\uHdb \RTbbinva \RTba $ implies that $ \uHda = \uB $, which reduces the left-hand side of~(\ref{equation_Riccati_6}) to 0, and implies $ \left( \RThatupaa \right)^{-1} = \left( \RTupaa \right)^{-1} $.

\section{Determinantal Expressions for $ {_0\!F_0} \left( \uS,\uLambda \right) $}
\label{section_uGamma1_distribution_Rician_Rayleigh_v}

\subsection{Expression for when Both $\uS$, $ \uLambda $ with Distinct Eigenvalues}
\label{section_distinct_eigenvalues}

Given $ \sigma_1 > \sigma_2 > \cdots > \sigma_{\NR} $ and $ \lambda_1 > \lambda_2 > \cdots > \lambda_{\NR} $, let us define
\begin{eqnarray}
\label{equation_g_sigma_lambda_distinct_eigenvalues}
g(\usigma, \ulambda) = g(\sigma_1, \cdots, \sigma_{\NR}, \lambda_1, \cdots, \lambda_{\NR}) = \frac{\det \left( e^{\sigma_i \lambda_j}\right)}{\prod_{i<j} (\sigma_i -\sigma_j) \prod_{i<j} (\lambda_i -\lambda_j) },
\end{eqnarray}
where $ \det \left( e^{\sigma_i \lambda_j} \right) $ is the determinant of the $ \NR \times \NR $ matrix with elements $ [\uD]_{i,j} = e^{\sigma_i \lambda_j} $, $ i, j = 1 : \NR $.

\begin{lemma}[{\cite{gross_jap_89}}{\cite{mckay_tcomm_09}}{\cite{chiani_tit_10}}]
If $ \NR \times \NR $  matrices $ \uS $ and $ \uLambda $ both have distinct eigenvalues, i.e., $ \sigma_1 > \sigma_2 > \cdots > \sigma_{\NR} $ and $ \lambda_1 > \lambda_2 > \cdots > \lambda_{\NR} $, then
\begin{eqnarray}
\label{equation_int_SUTU}
{_0\!F_0} \left( \uS,\uLambda \right) = g(\usigma, \ulambda) \phi(\NR),
\end{eqnarray}
where $ \phi(\NR) = \prod_{j = 1}^{\NR} (j-1)! $.
\end{lemma}

\subsection{New Expression for when Both $ \uS $, $ \uLambda $ May Have Non-Distinct Eigenvalues}
\label{section_nondistinct_nondistinct}

Let the distinct eigenvalues of $ \uS $ and $ \uLambda $ be ordered as follows
\begin{eqnarray}
\sigma^0_{(1)} > \sigma^0_{(2)} > \cdots > \sigma^0_{(m_{\cal{S}})},  \\
\lambda^0_{(1)} > \lambda^0_{(2)} > \cdots > \lambda^0_{(m'_{\cal{L}})}.
\end{eqnarray} 
The multiplicity of $ \sigma^0_{(i)} $ is denoted with $ m_i $, $ i = 1 : {\cal{S}} $.
The multiplicity of $ \lambda^0_{(i)} $ is denoted with $ m'_i $, $ i = 1 : {\cal{L}} $.
Let $ \usigma^0 $ be the vector with $ \sigma^0_{(1)} $, $ \sigma^0_{(2)} $, \ldots, $ \sigma^0_{(m_{\cal{S}})} $ repeated according to their multiplicities.
Let $ \ulambda^0 $ be the vector with $ \lambda^0_{(1)} $, $ \lambda^0_{(2)} $, \ldots, $ \lambda^0_{(m_{\cal{S}})} $ repeated according to their multiplicities.
Finally, define
\begin{eqnarray}
& a_i = m_1 - i, \quad & \text{for } 1 \le i \le m_1, \nonumber \\
& a_i = \sum_{p=1}^{k+1} m_p - i, \quad & \text{for } \sum_{p=1}^{k} m_p < i \le \sum_{p=1}^{k+1} m_p, \nonumber\\
& b_j = m'_1 - j, \quad & \text{for }  1 \le j \le m'_1, \nonumber\\
& b_j = \sum_{p=1}^{k+1} m'_p - j, \quad & \text{for }  \sum_{p=1}^{k} m'_p < j \le \sum_{p=1}^{k+1} m'_p.\nonumber
\end{eqnarray}

\begin{lemma}
\label{lemma_general_expression_0F0}
The continuous extension of $ g(\usigma, \ulambda) $ from~(\ref{equation_g_sigma_lambda_distinct_eigenvalues}) at $ (\usigma^0, \ulambda^0) $ helps express $ {_0\!F_0} \left( \uS,\uLambda \right) $ from~(\ref{equation_int_SUTU}), for $ \uS $ and $ \uLambda $ with arbitrary eigenvalues, as
\begin{eqnarray}
\label{equation_Takemura_general_expression}
\frac{  { \det  \left( \left. \frac{\partial^{a_i + b_j} ( e^{\sigma_i \lambda_j} )}{ {\partial \sigma_i}^{a_i} {\partial \lambda_j}^{b_j}} \right|_{\substack{\sigma_i = [\usigma^0]_i \\ \lambda_j =  [\ulambda^0]_j }} \right) } \frac{ \phi(\NR)}
{\prod_{i = 1}^{{\cal{S}}} \phi(m_i) \prod_{i = 1}^{{\cal{L}}} \phi(m'_i) } }
{\prod_{i<j}^{{\cal{S}}} (\sigma^0_{(i)} - \sigma^0_{(j)})^{m_i m_j}\prod_{i<j}^{{\cal{L}}} (\lambda^0_{(i)} - \lambda^0_{(j)})^{m'_i m'_j} }.
\end{eqnarray}
\end{lemma}

\IEEEproof{Follows by generalizing\cite[Lemma~2]{chiani_tit_10}.}

Expression~(\ref{equation_Takemura_general_expression}) reduces to  previously derived expressions:
\begin{itemize}
\item \cite[Eq.~(10)]{chiani_tit_10}, for both $ \uS $ and $ \uLambda $ with distinct eigenvalues --- see also~(\ref{equation_int_SUTU}).
\item \cite[Eq.~(16)]{chiani_tit_10}, for $ \uS $ with distinct eigenvalues and $ \uLambda $ with one subset of equal eigenvalues.
\item \cite[Eq.~(18)]{chiani_tit_10}, for $ \uS $ with distinct eigenvalues and $ \uLambda $ with one subset of zero eigenvalues.
\end{itemize}

\subsection{New Expression for when $ \uS $ is Rank-$ v $ with Distinct Nonzero Eigenvalues, and $ \uLambda $ is Rank-$ N_v $ Idempotent}
\label{section_S_nonfull_rank_L_idempotent}

\begin{corollary}
If $ \uS $ and $ \uLambda $ are $ \NR \times \NR $ matrices, $ \uS $ of rank $ v $ and with the nonzero distinct eigenvalues\footnote{To simplify writing, we change the notation for $ \sigma $.} $ \sigma_i $, $ i = 1 : v $, and $ \uLambda $ of rank $ N_v $ and idempotent, then $ {_0\!F_0} \left( \uS,\uLambda \right)  $ is given by
\begin{eqnarray}
\label{equation_denominator_rank_v_idempotent}
\frac{\Delta_1(N_v,\NR, \uS) }{\prod_{i = 1}^{v} \sigma_{i}^{\NR - v}
   \prod_{i<j}^{v} (\sigma_{i} - \sigma_{j}) } \frac{\phi(\NR)}{\phi(\NR - v) \phi(\NR - N_v)
\phi(N_v) }
\end{eqnarray}
where $ \Delta_1(N_v,\NR, \uS)  $ is the determinant of the $ \NR \times \NR $ matrix with (elementary-function) elements
\begin{eqnarray}
\label{equation_Delta_2}
\begin{cases}
e^{\sigma_i} \sigma_i^{N_v - j}, & \text{if } i \le v, j \le N_v \\
\sigma_i^{\NR - j}, & \text{if } i \le v, j > N_v \\
(N_v - j)! \binom{\NR - i}{N_v - j}, & \text{if } i > v, j \le N_v, \NR - i \ge N_v - j \\
0, & \text{if } i > v, j \le N_v, \NR - i < N_v - j \\
 (\NR - i)!, & \text{if } i > v, j > N_v, i = j \\
0, & \text{if } i > v, j > N_v, i \neq j.
\end{cases} \nonumber
\end{eqnarray}
\end{corollary}

\IEEEproof{Follows from~(\ref{equation_Takemura_general_expression}).}

Substituting~(\ref{equation_denominator_rank_v_idempotent}) into~(\ref{equation_uGamma1_mgf_uU_1_avg_1}) yields the first known expression (in terms of the determinant of a matrix whose entries are elementary functions) for the m.g.f.~of $ \uGamma_1 $, i.e., for the SC in the NCWD Gramian matrix $ \uW = \uHH \uH $ obtained from matrix $ \uH = ( \begin{matrix} \uHa & \uHb \end{matrix} )  $ with mean $ ( \begin{matrix} \uHda & \mzero \end{matrix} )  $.

\subsection{New Expression for when $ \uS $ is Rank-$ 1 $, and $ \uLambda $ is Rank-$ N $ Idempotent}
\label{section_0_F_0_v_1}

For $ v = 1 $, $ N_v $ reduces to $ \NR - \NT + 1 = N $, matrix $ \uM \doteq \NR \times v $ reduces to vector $ \umu \doteq \NR \times 1 $, and $ \uS $ can be written from~(\ref{equation_S_definition}) as follows\footnote{Here, we replace matrix symbol $ \uM $ with vector symbol $ \umu $ used in\cite{siriteanu_twc_13}.}:
\begin{eqnarray}
\label{equation_S_rank_1}
\uS = \frac{s \Gamma_{\text{s}} }{1-s \Gamma_{\text{s}}/\RTupaa }  \umu \umu^{\cal{H}} = \frac{s \overbrace{\Gamma_{\text{s}}/\RTupaa}^{\Gamma_{K,1}} }{1-s \Gamma_{\text{s}}/\RTupaa } \overbrace{\RTupaa \| \umu \|^2}^{=\alpha} \frac{\umu}{\| \umu \|} \frac{\umu^{\cal{H}}}{\| \umu \|}, 
\end{eqnarray}
i.e., $ \uS $ is rank-1 and with the nonzero eigenvalue given by
\begin{eqnarray}
\label{equation_sigma1}
\sigma_1 = \frac{s \Gamma_{K,1}}{1-s \Gamma_{K,1} } \alpha.
\end{eqnarray}

\begin{lemma}
\label{corollary_int_det_S_1_Lambda}
If $ \uS $ and $ \uLambda $ are $ \NR \times \NR $ matrices, $ \uS $ of rank $ 1 $ with nonzero eigenvalue $ \sigma_1 $, and $ \uLambda $ of rank $ N $ and idempotent, then $ {_0\!F_0} \left( \uS,\uLambda \right)  $ is given by
\begin{eqnarray}
\label{equation_int_etr_SUTU_S_r1_L_idem}
{_0\!F_0} \left( \uS,\uLambda \right) = \underbrace{\frac{(\NR - 1)! }{
\phi(N)  \phi(\NR - N) }}_{=A} \frac{ \Delta_2(N, \NR, \sigma_1)}{\sigma_1^{\NR - 1}},
\end{eqnarray}
where $ \Delta_2(N, \NR, \sigma_1) $ is the determinant of the $ \NR \times \NR $ matrix with (elementary-function) elements
\begin{eqnarray}
\label{equation_D3_S_1_L_idemp_num}
\begin{cases}
e^{\sigma_1} \sigma_1^{N - j}, & \text{if } i = 1, j \le N \\
\sigma_1^{\NR - j}, & \text{if } i = 1, j > N \\
(N - j)! \binom{\NR - i}{N - j},  & \text{if } i > 1, j \le N, \NR - i \ge N - j \\
0, & \text{if } i > 1, j \le N, \NR - i < N - j \\
 (\NR - i)!, & \text{if } i > 1, j > N, i = j \\
0, & \text{if } i > 1, j > N, i \neq j. 
\end{cases}
\end{eqnarray}
\end{lemma}

\IEEEproof{Follows from~(\ref{equation_denominator_rank_v_idempotent}).}

\end{appendices}

\footnotesize


\end{document}